\documentclass[footinbib,aps,prb,twocolumn,superscriptaddress,groupedaddress,a4paper]{revtex4}  % Uncomment for review/submission version
\usepackage{graphicx}
\usepackage{dcolumn}
\usepackage{bm}
\usepackage{amssymb}
\usepackage{amsmath}
\usepackage{subfigure}
\usepackage{color}
\usepackage{lscape}
\begin{document}

\widetext
\title{Electronic structure evolution in dilute carbide Ge$_{1-x}$C$_{x}$ alloys\\and implications for device applications}

%%%%%%%%%%%%%%%%%%%%%%%%%%%%%%%%%%%%%%%%
%%%% Authors, affiliations and date %%%%
%%%%%%%%%%%%%%%%%%%%%%%%%%%%%%%%%%%%%%%%

\author{Christopher A.~Broderick}
\email{c.broderick@umail.ucc.ie} % Email address must be kept before affiliation(s)
\affiliation{Tyndall National Institute, University College Cork, Lee Maltings, Dyke Parade, Cork T12 R5CP, Ireland}
\affiliation{Department of Physics, University College Cork, Cork T12 YN60, Ireland}

\author{Michael D.~Dunne}
\affiliation{Tyndall National Institute, University College Cork, Lee Maltings, Dyke Parade, Cork T12 R5CP, Ireland}
\affiliation{Department of Physics, University College Cork, Cork T12 YN60, Ireland}

\author{Daniel S.~P.~Tanner}
\affiliation{Tyndall National Institute, University College Cork, Lee Maltings, Dyke Parade, Cork T12 R5CP, Ireland}

\author{Eoin P.~O'Reilly}
\affiliation{Tyndall National Institute, University College Cork, Lee Maltings, Dyke Parade, Cork T12 R5CP, Ireland}
\affiliation{Department of Physics, University College Cork, Cork T12 YN60, Ireland}

\date{\today}

%%%%%%%%%%%%%%%%%%%%%%%%%%%%%%%%%%%%%%%%%%%%%
%%%% Abstract, keywords and PACS numbers %%%%
%%%%%%%%%%%%%%%%%%%%%%%%%%%%%%%%%%%%%%%%%%%%%

\begin{abstract}

We present a theoretical analysis of electronic structure evolution in the highly-mismatched dilute carbide group-IV alloy Ge$_{1-x}$C$_{x}$. For ordered alloy supercells, we demonstrate that C incorporation strongly perturbs the conduction band (CB) structure by driving hybridisation of $A_{1}$-symmetric linear combinations of Ge states lying close in energy to the CB edge. This leads, in the ultra-dilute limit, to the alloy CB edge being formed primarily of an $A_{1}$-symmetric linear combination of the L-point CB edge states of the Ge host matrix semiconductor. Our calculations describe the emergence of a ``quasi-direct'' alloy band gap, which retains a significant admixture of indirect Ge L-point CB edge character. We then analyse the evolution of the electronic structure of realistic (large, disordered) Ge$_{1-x}$C$_{x}$ alloy supercells for C compositions up to $x = 2$\%. We show that short-range alloy disorder introduces a distribution of localised states at energies below the Ge CB edge, with these states acquiring minimal direct ($\Gamma$) character. Our calculations demonstrate strong intrinsic inhomogeneous energy broadening of the CB edge Bloch character, driven by hybridisation between Ge host matrix and C-related localised states. The trends identified by our calculations are markedly different to those expected based on a recently proposed interpretation of the CB structure based on the band anti-crossing model. The implications of our findings for device applications are discussed.

\end{abstract}

% \keywords{0.0X}
% \pacs{0.0X}

\maketitle

%%%%%%%%%%%%%%%%%%%%%%
%%%% Introduction %%%%
%%%%%%%%%%%%%%%%%%%%%%

\section{Introduction}
\label{sec:introduction}

% General introduction

The indirect fundamental band gaps of the group-IV semiconductors silicon (Si) and germanium (Ge) lead to intrinsically inefficient emission and absorption of light, rendering these materials unsuitable for applications in (active) photonic or photovoltaic devices. At present, the development of Si photonics is limited by a lack of direct-gap materials which are both suitable for applications in semiconductor lasers and light-emitting diodes, and compatible with established complementary metal-oxide semiconductor (CMOS) fabrication and processing infrastructure. \cite{Zhou_LSAA_2015,Saito_SST_2016,Thomson_JO_2016} Similarly, direct-gap semiconductors having a fundamental band gap of $\approx 1$ eV and lattice constants commensurate with growth on Ge are required to facilitate the development of highly efficient multi-junction solar cells. \cite{Yamaguchi_SE_2005,Yamaguchi_SE_2008,Roucka_IEEEJPV_2016} To overcome this challenge, there has been a strong surge of interest in engineering the band structure of group-IV materials -- in particular Ge, via strain or alloying -- to produce semiconductors possessing a direct fundamental band gap. \cite{Geiger_FM_2015,Reboud_PCGC_2017} To date these efforts have centred on the application of tensile strain to Ge, \cite{Zhang_PRL_2009,Kurdia_APL_2010,Sanchez-Perez_PNAS_2011,Suess_NP_2013} and on tin- (Sn-) containing Ge$_{1-x}$Sn$_{x}$ alloys. \cite{Kouvetakis_ARMR_2006,Moontragoon_SST_2007,Soref_PTRSA_2014,Zaima_STAM_2015,Wirths_NP_2015,Margetis_APL_2018} Given the opportunities and challenges associated with direct-gap group-IV semiconductors, broader interest in related group-IV alloys containing lead \cite{Huang_PB_2014,Huang_JAC_2017,Alahmad_JEM_2018,Liu_JAC_2019,Broderick_GePb_2019} (Pb) and carbon \cite{Stephenson_JAP_2016,Stephenson_JEM_2016,Broderick_IEEENano_2018,Kirwan_SST_2019} (C) has begun to develop.

% Dilute carbide group-IV alloys

Initial interest in dilute carbide group-IV alloys originated over two decades ago, as a means to ameliorate issues related to the high levels of strain in Si$_{x}$Ge$_{1-x}$/Ge heterostructures. \cite{Osten_JCG_1995,Jain_SST_1995,Finkman_JAP_2001} Related theoretical analyses have focused on the impact of C incorporation on the structural, vibrational and transport properties of ternary dilute Si$_{y}$Ge$_{1-x-y}$C$_{x}$ alloys. \cite{Rucker_PRB_1996,Vaughan_PRB_2012} To date, there have been few theoretical investigations of the implications of dilute C incorporation on the electronic structure of group-IV materials. The recent establishment of novel epitaxial techniques to enable substitutional incorporation of C in Ge opens up the potential to develop electronic, photonic and photovoltaic devices based on dilute carbide Ge$_{1-x}$C$_{x}$ alloys. \cite{Stephenson_JAP_2016} However, previous investigations of Ge$_{1-x}$C$_{x}$ alloys have provided a range of qualitatively conflicting conclusions, including observations of strong band gap bowing, \cite{Okinaka_JCG_2003} a linear increase in band gap with increasing C composition $x$, \cite{Roe_JVSTB_1999} or the emergence of a direct band gap for a limited range of C compositions. \cite{Kolodzey_JCG_1995}

% Previous theoretical work

Several studies have highlighted the challenges associated with the growth of high quality substitutional Ge$_{1-x}$C$_{x}$ alloys: because C-C bonds are stronger than Ge-C bonds, there is a strong tendency during growth to form C-related defect clusters. \cite{Gall_PRB_2000} Consistent with this analysis, Park et al.~\cite{Park_JAP_2002} showed that it would be virtually impossible to achieve fully substitutional growth of Ge$_{1-x}$C$_{x}$ alloys by conventional molecular beam epitaxy (MBE) techniques. Recently, Stephenson et al.~\cite{Stephenson_C_2016} presented a route to overcome this problem, using a hybrid gas/solid-source MBE approach with tetrakis(germyl)methane (4GeMe) as the C source, a molecule that has one C atom bonded to four Ge atoms. This enabled growth of high quality Ge$_{1-x}$C$_{x}$ alloys having C compositions $x \approx 0.2$\% C. \cite{Stephenson_JAP_2016} These samples have been characterised using structural techniques and photo-modulated reflectance spectroscopy, \cite{Stephenson_JAP_2016} but there has to date been no reports of optical emission. Given the limited availability of experimental and theoretical data for Ge$_{1-x}$C$_{x}$ alloys, there is little information available in the literature regarding the alloy electronic structure. Recently, two theoretical analyses based on density functional theory (DFT) calculations have provided significant new insight into the Ge$_{1-x}$C$_{x}$ electronic structure. Stephenson et al.~\cite{Stephenson_JEM_2016} used hybrid functional DFT to compute the band structure of ordered Ge$_{N-1}$C$_{1}$ ($x = \frac{1}{N}$) alloy supercells containing $N \leq 128$ atoms ($x \geq 0.78$\%), and demonstrated that C incorporation strongly perturbs the conduction band (CB) structure. On the basis of these calculations the authors of Ref.~\onlinecite{Stephenson_JEM_2016} suggested that substitutional C acts as an isovalent impurity in Ge, giving rise to a C-related localised impurity state lying $\approx 0.4$ eV above the Ge conduction band (CB) minimum. It was further suggested that this C-related localised impurity state undergoes a band anti-crossing (BAC) interaction with the extended zone-centre $\Gamma_{7c}$ CB edge states of the Ge host matrix semiconductor -- similar to that in the III-V dilute nitride alloy GaN$_{x}$As$_{1-x}$ \cite{Shan_PRL_1999,Wu_SST_2002,Wu_PRB_2007,Reilly_SST_2009} -- resulting in (i) strong reduction of the fundamental band gap, (ii) the formation of a direct band gap for $x \gtrsim 0.8$\%, and (iii) closing of the alloy band gap for $x \lesssim 1.8$\%. However, suggestions regarding the presence and impact of a BAC interaction involving C-related localised states were drawn based on qualitative inspection of the calculated supercell band structures, without quantitative supporting analysis.

More recently, Kirwan et al.~\cite{Kirwan_SST_2019} also presented hybrid functional DFT calculations for ordered Ge$_{1-x}$C$_{x}$ alloy supercells. Here, supercell band structure calculations were supported by quantitative analysis of the character of the alloy CB edge states, as encapsulated in the band gap pressure coefficients. The calculated alloy band gap pressure coefficient was found to remain, independently of C composition $x$, approximately equal to that of the indirect (fundamental) L$_{6c}$-$\Gamma_{8v}$ band gap of the Ge host matrix, suggesting limited hybridisation of the CB edge with Ge $\Gamma_{7c}$ states. These calculations indicate the presence of a C-related localised state lying energetically within the band gap of the Ge host matrix, close in energy to the Ge CB minimum. However, analysis of the calculated alloy band gap pressure coefficients produced results inconsistent with the presence of a BAC interaction in Ge$_{1-x}$C$_{x}$, indicating instead that (i) C incorporation primarily drives hybridisation between the $\Gamma_{7c}$ and X$_{5c}$ CB edge states of Ge, and (ii) the alloy CB edge retains primarily indirect Ge L$_{6c}$ character.

% What we do/show in this paper

Given these conflicting reports, further theoretical insight is required to quantify the nature and evolution of the Ge$_{1-x}$C$_{x}$ electronic structure, so that the potential of the alloy for practical applications can be assessed. In this paper, we calculate the electronic structure of idealised (ordered) Ge$_{N-1}$C$_{1}$ alloy supercells, and of realistic (large, disordered) Ge$_{1728-M}$C$_{M}$ ($x = \frac{M}{1728}$) alloy supercells containing a statistically random distribution of $M$ substitutional C atoms. While previous DFT-based analyses have been limited to ordered supercells containing $\leq 128$ atoms ($x \geq 0.78$\%), we adopt a semi-empirical theoretical framework that allows high-throughput calculations to be performed for large supercells containing $\gtrsim 10^{3}$ atoms. Using this approach we quantify (i) the impact of C incorporation on the Ge electronic structure in the ultra-dilute (impurity) limit, (ii) the impact of short-range alloy disorder (including C clustering) on the electronic structure, and (iii) the evolution of the electronic structure with C composition $x$ in large, disordered alloy supercells.

We explicitly demonstrate the presence of C-induced hybridisation of Ge host matrix CB edge states in ordered alloy supercells, and demonstrate that the Ge$_{1-x}$C$_{x}$ alloy CB edge retains primarily indirect (Ge L$_{6c}$) character. Explicit analysis of the CB eigenstates in supercells containing up to 2000 atoms provides a broader picture of the electronic structure evolution. Specifically, we confirm that Ge$_{1-x}$C$_{x}$ admits a ``quasi-direct'' band gap in ordered alloy supercells: while the CB minimum appears at the zone centre of a Ge$_{N-1}$C$_{1}$ supercell, the associated eigenstate in selected supercells is formed predominantly of a linear combination of Ge L$_{6c}$ CB edge states having purely $s$-like orbital character at the C lattice site. We further demonstrate that the alloy CB edge exhibits minimal localisation with increasing supercell size, challenging the suggestion that the introduction of an isolated substitutional C atom in Ge generates a (strongly) localised impurity state. For disordered alloy supercells we find that short-range alloy disorder gives rise to a distribution of C-related localised states -- associated with nearest-neighbour C-C pairs, as well as larger clusters of substitutional C atoms and various C-Ge-C type neighbour complexes -- lying energetically within the Ge band gap, with these states acquiring minimal direct (Ge $\Gamma_{7c}$) character. Overall, our analysis reveals behaviour that is markedly different to that expected on the basis of the BAC model and, in agreement with the conclusions of Kirwan et al., \cite{Kirwan_SST_2019} we demonstrate that C incorporation does not drive the formation of a direct band gap.

% organization of this paper

The remainder of this paper is organized as follows. In Sec.~\ref{sec:theoretical_model} we describe the semi-empirical framework we have established to calculate the structural and electronic properties of Ge$_{1-x}$C$_{x}$ alloys. The results of our calculations are presented in Sec.~\ref{sec:results}, beginning in Sec.~\ref{sec:results_ordered} with an analysis of the impact of C incorporation on the band structure of small ordered Ge$_{1-x}$C$_{x}$ alloy supercells. This is followed by a respective analysis of the character -- in Sec.~\ref{sec:results_impurity_limit} -- and localisation -- in Sec.~\ref{sec:results_localisation} -- of the CB states as a function of $x$ in ordered supercells containing up to $N = 2000$ atoms. In Sec.~\ref{sec:results_disordered} we analyse the electronic structure evolution in large, disordered Ge$_{1-x}$C$_{x}$ alloy supercells. In Sec.~\ref{sec:implications} the implications of our results for device applications are briefly described. Finally, in Sec.~\ref{sec:conclusions} we summarise and conclude.

%%%%%%%%%%%%%%%%%%%%%%%%%%%
%%%% Theoretical model %%%%
%%%%%%%%%%%%%%%%%%%%%%%%%%%

\section{Theoretical model}
\label{sec:theoretical_model}

% Choice of methods

Highly-mismatched semiconductor alloys are characterised by constituent elements differing significantly in size (covalent radius) and chemical properties (valence orbital energies), resulting in (i) significant changes to the electronic structure of the host matrix semiconductor in response to incorporation of dilute concentrations of the alloying element, and (ii) strong sensitivity of the electronic structure to short-range alloy disorder. \cite{Wu_SST_2002,Wu_PRB_2007,Reilly_SST_2009} From a technical perspective, these factors constitute a breakdown of the virtual crystal approximation (VCA), mandating direct atomistic calculations in order to capture -- even qualitatively -- the evolution of the alloy electronic structure.

Beginning with an $N$-atom Ge$_{N}$ supercell, the smallest C composition that can be treated is $x = \frac{1}{N}$ in an ordered Ge$_{N-1}$C$_{1}$ alloy supercell. Investigation of the material properties in the dilute composition (impurity) limit is therefore limited by the maximum supercell size that can be treated by the electronic structure method employed, and the use of ordered supercells to achieve dilute compositions precludes the investigation of alloy disorder effects. First principles methods that accurately describe the electronic properties of semiconductors are in practice limited to supercells containing $N \lesssim 10^{2}$ atoms, allowing access to compositions $\sim 1$\% but providing limited scope to investigate isolated impurities or to quantify alloy disorder effects.

Our theoretical analysis of dilute Ge$_{1-x}$C$_{x}$ alloys is therefore based on a semi-empirical framework, which can be extended to large system sizes to enable quantitative atomistic analysis of the impact of alloy disorder and related effects in realistic ($N \gtrsim 10^{3}$, disordered) alloy supercells. Firstly, we use a parametrised valence force field (VFF) potential to perform structural relaxation of alloy supercells. \cite{Tanner_thesis_2017} Secondly, the electronic structure of relaxed alloy supercells is computed using a nearest-neighbour $sp^{3}s^{\ast}$ tight-binding (TB) Hamiltonian. \cite{Broderick_IEEENano_2018} Both the VFF potential and TB Hamiltonian are parametrised via the bulk structural, elastic and electronic properties of the constituent materials -- i.e.~the elemental and compound materials formed by nearest-neighbour bonds in a given alloy supercell -- which for Ge$_{1-x}$C$_{x}$ are the elemental diamond-structured group-IV semiconductors Ge and C, and the zinc blende IV-IV compound GeC (zb-GeC). \cite{Tanner_elastic_2019,Broderick_electronic_2019}

% Valence force field potential

\subsection{Valence force field potential}
\label{sec:theoretical_model_vff}

We use the modified form \cite{Martin_PRB_1970} of the VFF potential introduced by Musgrave and Pople, \cite{Musgrave_PRSLA_1962} whereby the contribution to the lattice free energy associated with an atom located at lattice site $i$ is

\begin{align}
	V_{i} &= \frac{1}{2} \sum_{j} \frac{ k_{r} }{2} \left( r_{ij} - r_{ij}^{(0)} \right)^{2} \nonumber \\
	&+ \sum_{j} \sum_{k > j} \bigg[ \frac{ k_{\theta} }{2} r_{ij}^{(0)} r_{ik}^{(0)} \left( \theta_{ijk} - \theta_{ijk}^{(0)} \right)^{2} \nonumber \\
	&\hspace{1.5cm}+ k_{rr} \left( r_{ij} - r_{ij}^{(0)} \right) \left( r_{ik} - r_{ik}^{(0)} \right) \nonumber \\
	&+ k_{r\theta} \bigg( r_{ij}^{(0)} \left( r_{ij} - r_{ij}^{(0)} \right) + r_{ik}^{(0)} \left( r_{ik} - r_{ik}^{(0)} \right) \bigg) \left( \theta_{ijk} - \theta_{ijk}^{(0)} \right) \bigg] \, ,
	\label{eq:vff_potential}
\end{align}

\noindent
where $j$ and $k$ index the nearest-neighbour atoms of atom $i$, $r_{ij}^{(0)}$ and $r_{ij}$ respectively denote the unstrained (equilibrium) and relaxed bond lengths between atoms $i$ and $j$, and $\theta_{ijk}^{(0)}$ and $\theta_{ijk}$ respectively denote the unstrained and relaxed angles formed by adjacent nearest neighbour bonds, between atoms $i$ and $j$, and between atoms $i$ and $k$. The first and second terms in Eq.~\eqref{eq:vff_potential} respectively describe contributions to the lattice free energy associated with pure bond stretching and pure bond-angle bending, while the third and fourth terms are ``cross terms'' which respectively describe the impact of changes in $r_{ik}$ on $r_{ij}$, and the impact of changes in $\theta_{ijk}$ on both $r_{ij}$ and $r_{ik}$.

Re-casting Eq.~\eqref{eq:vff_potential} in terms of macroscopic and internal strains allows the force constants $k_{r}$, $k_{\theta}$, $k_{rr}$ and $k_{r\theta}$ to be determined analytically in terms of the elastic constants $C_{11}$, $C_{12}$ and $C_{44}$, and the Kleinman (internal strain) parameter $\zeta$ for diamond or weakly-polar zinc blende structured materials. \cite{Tanner_thesis_2017,Tanner_PRB_2019} This provides an exact description of the static lattice properties in the linear elastic limit, circumventing the conventional requirement to determine the VFF force constants via numerical fitting. The unstrained bond lengths and VFF force constants used in our calculations are provided in Table~\ref{tab:vff_parameters}. Full details of the parametrisation (via hybrid functional DFT calculations) and benchmarking (vs.~hybrid functional DFT alloy supercells relaxations) of Eq.~\eqref{eq:vff_potential} for Ge$_{1-x}$C$_{x}$ and related group-IV alloys will be presented in Ref.~\onlinecite{Tanner_elastic_2019}. Structural relaxations -- implemented using the General Utility Lattice Program (\textsc{GULP}) \cite{Gale_JCSFT_1997,Gale_MS_2003,Gale_ZK_2005} -- for Ge$_{1-x}$C$_{x}$ alloy supercells proceed by minimising the lattice free energy computed via Eq.~\eqref{eq:vff_potential}, by allowing the supercell lattice vectors and ionic positions to relax freely.

% Tight-binding Hamiltonian

\subsection{Tight-binding Hamiltonian}
\label{sec:theoretical_model_tb}

% Table 1

\begin{table}[t!]
	\caption{\label{tab:vff_parameters} Equilibrium bond lengths $r^{(0)}$, and force constants $k_{r}$, $k_{\theta}$, $k_{rr}$ and $k_{r\theta}$, used to implement structural relaxations for Ge$_{1-x}$C$_{x}$ alloy supercells using the VFF potential of Eq.~\eqref{eq:vff_potential}. Force constants have been computed analytically based on DFT-calculated structural properties for Ge, C and zb-GeC. \cite{Tanner_thesis_2017,Tanner_elastic_2019}}
	\begin{ruledtabular}
		\begin{tabular}{clccc}
			Parameter     & Unit                     & Ge     & C           & zb-GeC   \\
			\hline
			$r^{(0)}$     & \AA                      & 2.445  &   1.530     &  1.969   \\
			$k_{r}$       & eV \AA$^{-2}$            & 7.0414 &  26.4077    & 11.8086  \\
			$k_{\theta}$  & eV \AA$^{-2}$ rad$^{-2}$ & 0.5104 &   3.7208    &  1.0780  \\
			$k_{rr}$      & eV \AA$^{-2}$            & 0.2416 &   0.9740    &  1.0557  \\
			$k_{r\theta}$ & eV \AA$^{-2}$ rad$^{-1}$ & 0.3005 &   1.7832    &  1.2515  \\
		\end{tabular}
	\end{ruledtabular}
\end{table}

% Figure 1

\begin{figure*}[t!]
	\includegraphics[width=1.00\textwidth]{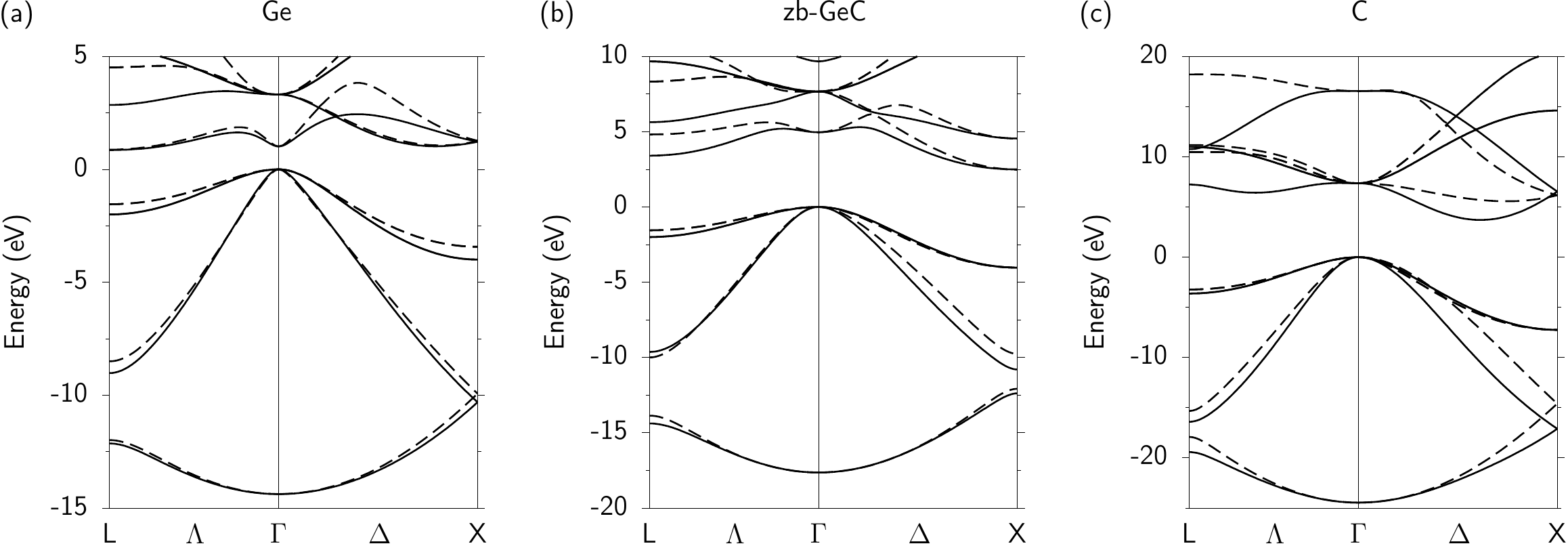}
	\caption{Band structure of (a) Ge, (b) zb-GeC, and (c) C, calculated via DFT using the HSEsol exchange-correlation functional (dashed lines), and via a semi-empirical $sp^{3}s^{\ast}$ TB Hamiltonian (solid lines). All calculations omit spin-orbit coupling. For comparative purposes, the zero of energy has been chosen to lie at the Fermi level (VB edge) in all cases. Note the differences in scales on the ordinates of (a), (b) and (c).}
	\label{fig:primitive_band_structures}
\end{figure*}

Given its use of a localised basis of atomic orbitals, the TB method is well suited to analyse the impact of localised impurities on the electronic structure of a given host matrix semiconductor. \cite{Reilly_JPCS_2010} This is of particular importance for highly-mismatched alloys -- such as Ge$_{1-x}$C$_{x}$ -- in which the constituent elements have significant differences in size and chemical properties. We employ a nearest-neighbour $sp^{3}s^{\ast}$ TB Hamiltonian, \cite{Vogl_JPCS_1983} based closely on that we have recently established for Ge$_{1-x}$Sn$_{x}$ alloys. \cite{Halloran_OQE_2019} We note however that the TB model employed here differs from that of Ref.~\onlinecite{Halloran_OQE_2019} in that we omit spin-orbit coupling. This is justified on the basis that DFT calculations have demonstrated that C incorporation in Ge primarily impacts the band structure close in energy to the Ge CB edge, \cite{Stephenson_JAP_2016,Kirwan_SST_2019} where spin-orbit coupling has minimal impact. Henceforth, we therefore refer to high-symmetry eigenstates of Ge using the conventional $O_{h}$ ($m\bar{3}m$) point group notation for the diamond lattice. \cite{Dresselhaus_book_2006}

It is well established that the accuracy of the TB fit to the band structure of a given semiconductor material can be improved via the inclusion of $d$-like atomic orbitals, to obtain a $sp^{3}s^{\ast}d^{5}$ basis set. \cite{Jancu_PRB_1998} We have chosen to employ an $sp^{3}s^{\ast}$ rather than $sp^{3}s^{\ast}d^{5}$ basis in our calculations for two reasons. Firstly, while the $sp^{3}s^{\ast}d^{5}$ basis allows for more accurate fitting of a given target band dispersion, this comes at the cost of doubling the size of the Hamiltonian for a supercell containing a given number of atoms compared to that associated with a $sp^{3}s^{\ast}$ basis. For the large supercell sizes required to simulate the properties of realistic highly-mismatched alloys, this doubling of the size of the supercell Hamiltonian represents a significant increase in the computational cost of numerical diagonalisation. Secondly, while the $sp^{3}s^{\ast}d^{5}$ basis enables a more accurate fit to a range of band-edge effective masses and higher energy CB states, the $sp^{3}s^{\ast}$ basis is sufficient to accurately describe the band energies at high-symmetry points in the Brillouin zone. Since our aim here is to describe the nature of the alloy band gap, we are interested primarily in C-induced band hybridisation effects. The strength with which different eigenstates of a given semiconductor hybridise in response to a perturbation generally depends critically on the separation in energy of a small number of high-symmetry states. For dilute Ge$_{1-x}$C$_{x}$ alloys we are interested in identifying the potential presence of an indirect- to direct-gap transition, so it is crucial that the energy of the L$_{1c}$ (L-point CB minimum) states are described accurately relative to the $\Gamma_{2'c}$ (zone-centre CB edge) state. For the purpose of describing the nature and evolution of the band gap of a large number of semiconductor alloys the $sp^{3}s^{\ast}$ basis has been found to be sufficient, representing a small trade-off in accuracy in order to significantly reduce parametric complexity and computational cost. Indeed, we have previously demonstrated that a TB Hamiltonian employing an $sp^{3}s^{\ast}$ basis provides quantitatively accurate insight into the properties of highly-mismatched III-V semiconductor alloys containing nitrogen \cite{Lindsay_SSC_1999,Reilly_SST_2002,Lindsay_PRL_2004} (N), boron \cite{Lindsay_PRB_2007,Lindsay_PSSC_2008,Sander_PRB_2011} (B) or bismuth \cite{Usman_PRB_2011,Usman_PRB_2013,Broderick_PRB_2014,Usman_PRA_2018} (Bi).

The impact of lattice relaxation (local strain) is incorporated in the TB Hamiltonian via bond length- and angle-dependent inter-atomic interaction matrix elements using, respectively, the generalised form of Harrison's rule \cite{Brey_PRB_1984,Priester_PRB_1988} and the Slater-Koster two-centre integrals. \cite{Slater_PR_1954} To overcome the failure of this conventional parametrisation to describe deformation potentials associated with tetragonal (biaxial) deformations, we include two additional strain-related terms. Firstly, we include an on-site correction of the $p$ orbital energies, which provides an accurate description of the $\Gamma$-point valence band (VB) edge axial deformation potential $b$. \cite{Rucker_PSSB_1983} Secondly, we include a correction to the $V_{s^{\ast}p\sigma}$ inter-atomic interaction matrix elements, which accounts for the influence of $d$ orbital interactions in determining the axial deformation potential $\Xi_{u}^{\text{X}}$ associated with the X-point CB edge states. \cite{Munoz_PRB_1993} To incorporate these corrections in alloy supercell calculations we have cast them in local form, by writing the infinitesimal strain tensor at each lattice site in terms of the relaxed nearest-neighbour bond lengths and angles about that site. For a given relaxed alloy supercell, the construction of the supercell Hamiltonian -- which accounts explicitly for size and chemical differences between constituent elements -- proceeds as described in Ref.~\onlinecite{Halloran_OQE_2019}.

Our TB parameters are obtained by fitting to selected high-symmetry point energies determined via hybrid functional (HSEsol) DFT calculations. \cite{Broderick_electronic_2019} Since C incorporation in Ge primarily impacts the band structure close in energy to the CB edge of the Ge host matrix semiconductor, our priority in fitting TB parameters for Ge was to describe the energies of the $\Gamma_{2'c}$, L$_{1c}$ and X$_{1c}$ high-symmetry $\Gamma$-, L- and X-point CB edge states. \cite{Halloran_OQE_2019} For C and zb-GeC we follow the fitting procedure outlined by Vogl et al.~\cite{Vogl_JPCS_1983} without modification. Full details of the TB parametrisation for Ge$_{1-x}$C$_{x}$ and related group-IV alloys will be presented in Ref.~\onlinecite{Broderick_electronic_2019}. The zero of energy for our alloy TB Hamiltonian is set at the Ge $\Gamma_{25'v}$ VB edge. We assume a natural VB offset of $-4.24$ eV between C and Ge, following the first principles calculations of Li et al. \cite{Li_APL_2009} Based on a linear interpolation of this VB offset with respect to the HSEsol-calculated lattice constants of C and Ge, \cite{Tanner_elastic_2019} we estimate a VB offset of $-2.12$ eV between Ge and zb-GeC.

The resulting TB fits to the Ge, zb-GeC and C band structures are shown respectively in Figs.~\ref{fig:primitive_band_structures}(a),~\ref{fig:primitive_band_structures}(b) and~\ref{fig:primitive_band_structures}(c), where solid (dashed) black lines show the TB-calculated (reference HSEsol DFT \cite{Broderick_electronic_2019}) band structure. Note the differences in scales on the ordinates of Figs.~\ref{fig:primitive_band_structures}(a),~\ref{fig:primitive_band_structures}(b) and~\ref{fig:primitive_band_structures}(c). For zb-GeC and C the underestimation of the L-point CB edge energies represents a typical $sp^{3}s^{\ast}$ fit. \cite{Vogl_JPCS_1983} We note however that these discrepancies, which are largest in our TB fit to the band structure of C, should have minimal impact in alloy supercell calculations. Since we employ a nearest-neighbour TB Hamiltonian, parameters for C are only employed in the construction of the supercell Hamiltonian when C atoms appear as nearest neighbours. In a randomly disordered substitutional Ge$_{1-x}$C$_{x}$ alloy having C composition $x$, the probability for small $x$ of two C atoms occupying nearest-neighbour lattice sites is $2 x^{2}$ -- i.e.~a randomly disordered $N$-atom Ge$_{1-x}$C$_{x}$ supercell will contain, on average, a total of $N \times 2 x^{2}$ C-C nearest-neighbour pairs. Since we are concerned only with dilute C compositions $x \lesssim 2$\%, for the supercells considered in our analysis -- which contain $N \leq 2000$ atoms -- we expect $< 2$ C-C pairs to be present in any given disordered alloy supercell. This indeed turns out to be the case for the disordered supercells considered in Sec.~\ref{sec:results_disordered}, in which the C atoms are substituted at statistically randomly selected lattice sites. In addition, as we describe in Sec.~\ref{sec:results_ordered}, the results of our TB supercell calculations for small, ordered Ge$_{1-x}$C$_{x}$ supercells are in good quantitative agreement with hybrid functional DFT calculations, \cite{Kirwan_IEEENano_2018,Kirwan_SST_2019} confirming the validity of our TB model to investigate Ge$_{1-x}$C$_{x}$ alloys.

%%%%%%%%%%%%%%%%%
%%%% Results %%%%
%%%%%%%%%%%%%%%%%

\section{Results}
\label{sec:results}

In this section we present the results of our analysis of the electronic structure of dilute Ge$_{1-x}$C$_{x}$. We begin in Sec.~\ref{sec:results_ordered} by considering the impact of C incorporation on the band structure of ordered alloy supercells, and then in Secs.~\ref{sec:results_impurity_limit} and~\ref{sec:results_localisation} we respectively analyse in detail the character and localisation of the alloy CB edge states as a function of C composition $x$. Next, in Sec.~\ref{sec:results_disordered} we turn our attention to the evolution of the electronic structure with $x$ in realistic (large, disordered) alloy supercells.

Since the Ge$_{1-x}$C$_{x}$ band edge evolves from the L$_{1c}$ CB minimum states of Ge, and since we are interested in the potential evolution of a direct band gap in Ge$_{1-x}$C$_{x}$ alloys -- whereby the alloy CB edge would acquire predominately Ge $\Gamma_{2'c}$ character -- we restrict our attention primarily to supercells in which the L-point states of the underlying diamond structure fold to the supercell zone centre $\textbf{K} = 0$ -- i.e.~$n \times n \times n$ face-centred cubic (FCC) or simple cubic (SC) supercells for even values of $n$. This allows for C-induced hybridisation of the L$_{1c}$ CB minimum and $\Gamma_{2'c}$ zone-centre CB edge states of Ge, which can be expected to be important given the small $\approx 0.15$ eV separation in energy between the fundamental (indirect) and direct band gaps of Ge. Indeed, we have recently demonstrated the importance of such effects in determining the nature of the indirect- to direct-gap transition in Ge$_{1-x}$Sn$_{x}$ and Ge$_{1-x}$Pb$_{x}$ alloys. \cite{Halloran_OQE_2019,Eales_SR_2019,Broderick_GePb_2019,Broderick_NUSOD_2019} Our calculations suggest, in agreement with hybrid functional DFT calculations, \cite{Stephenson_JAP_2016,Kirwan_SST_2019} that C incorporation has minimal impact on the VB structure, so we focus our analysis on the alloy CB structure.

% Figure 2

\begin{figure*}[ht!]
	\includegraphics[width=1.00\textwidth]{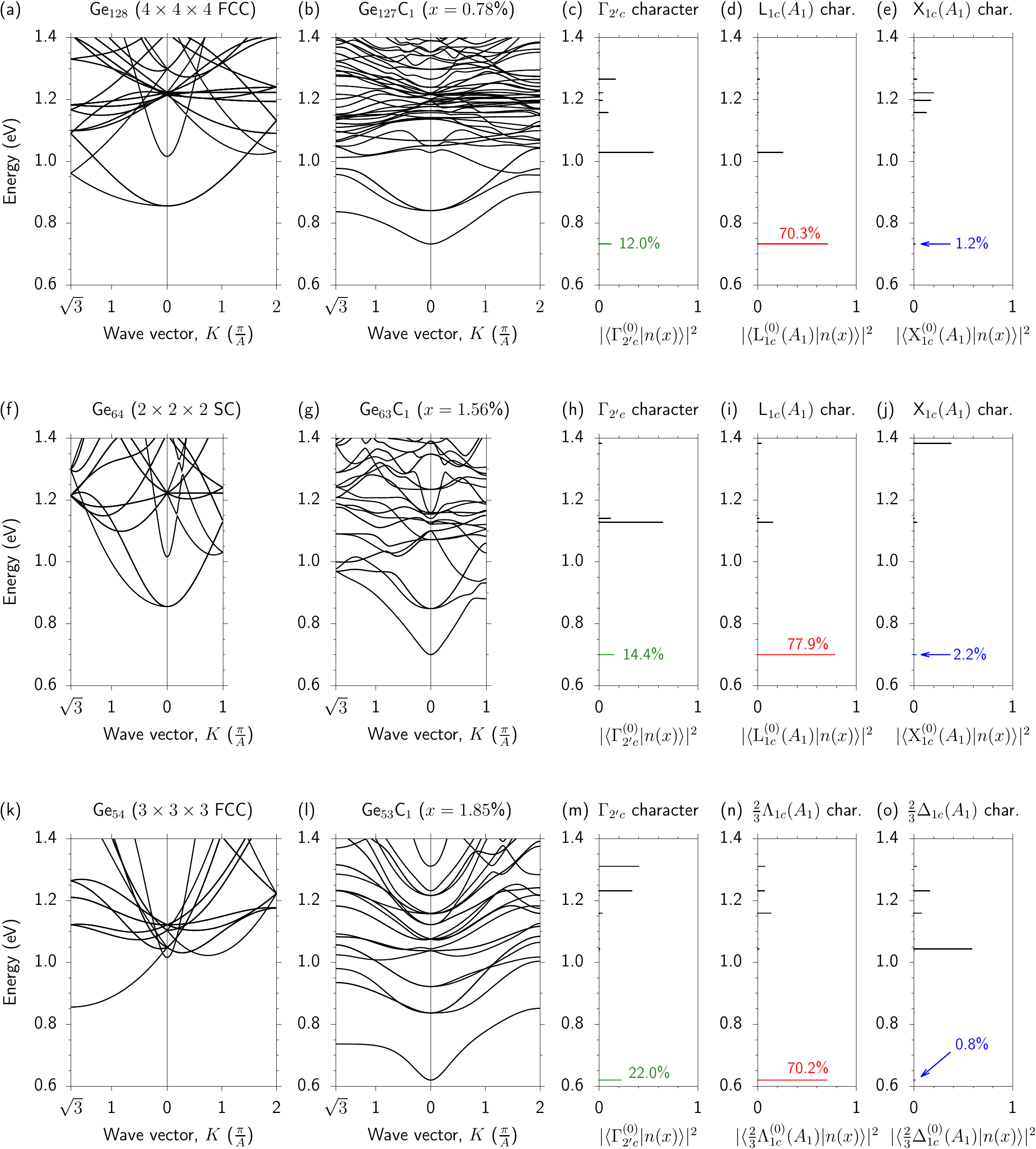}
	\caption{Top row: Calculated CB structure for (a) a Ge$_{128}$ supercell, and (b) an ordered Ge$_{127}$C$_{1}$ ($x = 0.78$\%) supercell. (c) -- (e) respectively show the calculated fractional Ge $\Gamma_{2'c}$, L$_{1c} ( A_{1} )$ and X$_{1c} ( A_{1} )$ character spectra of the Ge$_{127}$C$_{1}$ supercell of (b). Middle row: Calculated CB structure for (f) a Ge$_{64}$ supercell, and (g) an ordered Ge$_{63}$C$_{1}$ ($x = 1.56$\%) supercell. (h) -- (j) respectively show the calculated fractional Ge $\Gamma_{2'c}$, L$_{1c} ( A_{1} )$ and X$_{1c} ( A_{1} )$ character spectra for the Ge$_{63}$C$_{1}$ supercell of (g). Bottom row: Calculated CB structure for (k) a Ge$_{54}$ supercell, and (l) an ordered Ge$_{53}$C$_{1}$ ($x = 1.85$\%) alloy supercell. (m) -- (o) respectively show the calculated fractional Ge $\Gamma_{2'c}$, $\frac{2}{3} \Lambda_{1c} ( A_{1} )$ and $\frac{2}{3} \Delta_{1c} ( A_{1} )$ character spectrum for the Ge$_{53}$C$_{1}$ supercell of (l). The zero of energy of all band structure plots is set at the Ge VB edge. Green, red and blue coloring is used to highlight the Ge $\Gamma_{2'c}$, L$_{1c} ( A_{1} )$ (or $\frac{2}{3} \Lambda_{1c} ( A_{1} )$), and X$_{1c} ( A_{1} )$ (or $\frac{2}{3} \Delta_{1c} ( A_{1} )$) character of the CB edge state in each alloy supercell.}
	\label{fig:ordered_supercells_bands}
\end{figure*}

%%%%%%%%%%%%%%%%%%%%%%%%%%%%%%%%%%%%%%%%%%%%%%%%%%%%%%%%%%%%%%%%%%%%%%%%%%%%%%%
%%%% Results: Band structure of ordered Ge$_{1-x}$C$_{x}$ alloy supercells %%%%
%%%%%%%%%%%%%%%%%%%%%%%%%%%%%%%%%%%%%%%%%%%%%%%%%%%%%%%%%%%%%%%%%%%%%%%%%%%%%%%

\subsection{Band structure of ordered Ge$_{1-x}$C$_{x}$ alloy supercells}
\label{sec:results_ordered}

% Ge(127)C(1) band structure

Figure~\ref{fig:ordered_supercells_bands}(b) shows the calculated CB structure of an ordered, 128-atom Ge$_{127}$C$_{1}$ ($x = 0.78$\%) supercell. For comparative purposes, the CB structure of the corresponding C-free Ge$_{128}$ supercell is shown in Fig.~\ref{fig:ordered_supercells_bands}(a). The band dispersion is plotted as a function of the supercell wave vector $\textbf{K}$ in units of $\frac{ \pi }{ A }$, where $A$ is the supercell lattice constant, equal respectively to $\frac{ na }{ 2 }$ or $na$ for an $n \times n \times n$ FCC or SC supercell. The lowest energy CB states at $\textbf{K} = 0$ in Ge$_{128}$ are the folded L$_{1c}$ CB edge states, with the $\Gamma_{2'c}$ zone-centre state lying 160 meV higher in energy. For the Ge$_{127}$C$_{1}$ supercell we firstly note that C incorporation strongly perturbs the CB structure, leading to a large band gap reduction of 91 meV compared to the fundamental band gap of Ge ($E_{g} = 0.856$ eV, in the absence of spin-orbit coupling). Additionally, we note that the alloy CB minimum lies at $\textbf{K} = 0$. Indeed, based on qualitative inspection of Fig.~\ref{fig:ordered_supercells_bands}(b) it is tempting to conclude -- given the strong band gap reduction and apparent emergence of a C-related impurity band lying energetically within the Ge band gap -- that C acts as an isovalent impurity, driving strong band gap reduction via a BAC interaction, as suggested by Stephenson et al. \cite{Stephenson_JEM_2016,Stephenson_JAP_2016} However, we note that the CB edge state is non-degenerate, while the second lowest energy set of CB states -- lying 108 meV above the CB edge in energy -- is threefold degenerate. This suggests C-induced splitting of the fourfold degenerate L$_{1c}$ CB edge states of Ge, and that the alloy CB edge might be better described in terms of a linear combination of Ge L$_{1c}$ states, rather than as a C-related localised impurity state.

To ascertain whether or not this is the case, we have undertaken a quantitative analysis of the character of the alloy CB states. Figure~\ref{fig:ordered_supercells_bands}(c) shows the fractional Ge $\Gamma_{2'c}$ character of the Ge$_{127}$C$_{1}$ CB states, calculated by projecting the $\Gamma_{2'c}$ eigenstate $\vert \Gamma_{2'c}^{(0)} \rangle$ of the Ge$_{128}$ host matrix supercell onto the full spectrum $\lbrace \vert n (x) \rangle \rbrace$ of $\textbf{K} = 0$ Ge$_{127}$C$_{1}$ alloy CB eigenstates. Note that we use the superscript ``(0)'' henceforth to denote unperturbed Ge host matrix eigenstates. We calculate that the Ge$_{127}$C$_{1}$ CB edge eigenstate -- highlighted in Fig.~\ref{fig:ordered_supercells_bands}(c) via green coloring -- acquires a small (12.0\%) admixture of Ge $\Gamma_{2'c}$ character. As a consequence of their symmetry the second lowest energy set of CB states acquire no Ge $\Gamma_{2'c}$ character, while the third lowest energy state -- originating from the Ge $\Gamma_{2'c}$ state and lying 296 meV above the alloy CB edge in energy -- retains majority (54.4\%) Ge $\Gamma_{2'c}$ character. We find that the remainder of the Ge $\Gamma_{2'c}$ character is spread over a small number of higher energy alloy CB states, originating primarily from the folded $X_{1c}$ CB edge states of Ge. Our calculations therefore suggest that the Ge$_{127}$C$_{1}$ CB edge acquires only minimal direct (Ge $\Gamma_{2'c}$) character. To confirm this we calculate the pressure coefficient $\frac{ \textrm{d} E_{g} }{ \textrm{d} P }$ associated with the fundamental band gap. We have demonstrated elsewhere \cite{Halloran_OQE_2019,Eales_SR_2019,Broderick_GePb_2019} that calculation of band structure as a function of hydrostatic pressure can provide useful quantitative insight into the character of the band edge states, and hence the nature and evolution of the alloy band gap. The pressure coefficients associated the indirect L$_{1c}$-$\Gamma_{25'v}$, direct $\Gamma_{2'c}$-$\Gamma_{25'v}$ and indirect X$_{1c}$-$\Gamma_{25'v}$ band gaps of Ge are significantly different to on another, having respective values $\frac{ \textrm{d} E_{g} }{ \textrm{d} P } = 13.33$, 4.66 and $-$1.60 meV kbar$^{-1}$ in the HSEsol DFT calculations to which we fit our TB parameters. \cite{Halloran_OQE_2019} Since we are dealing with dilute C compositions, we expect that an alloy having primarily indirect character will have a significantly lower pressure coefficient than an alloy having direct-gap character. We calculate $\frac{ \textrm{d} E_{g} }{ \textrm{d} P } = 5.25$ meV kbar$^{-1}$ for the Ge$_{127}$C$_{1}$ supercell confirming that, despite strong perturbation of the CB structure, the alloy CB edge retains primarily Ge L$_{1c}$ character. Our calculated value of $\frac{ \textrm{d} E_{g} }{ \textrm{d} P }$ here is in good agreement with the value of 4.55 meV kbar$^{-1}$ obtained from the hybrid functional DFT calculations of Ref.~\onlinecite{Kirwan_SST_2019}.

Direct inspection of the Ge$_{127}$C$_{1}$ CB edge eigenstate reveals purely $s$-like orbital character ($A_{1}$ symmetry) at the C lattice site. Since, in general, alloying drives hybridisation between host matrix states having the same symmetry, we conclude that C incorporation drives hybridisation primarily between $\vert \Gamma_{2'c}^{(0)} \rangle$ and a linear combination $\vert \textrm{L}_{1c}^{(0)} ( A_{1} ) \rangle$ of Ge L$_{1c}$ eigenstates having purely $s$-like orbital character at the C lattice site. We confirm that this is the case by using the $\textbf{K} = 0$ eigenstates of a C-free Ge$_{128}$ supercell to explicitly construct $\vert \textrm{L}_{1c}^{(0)} ( A_{1} ) \rangle$, and then calculate the corresponding Ge L$_{1c} ( A_{1} )$ character of the Ge$_{127}$C$_{1}$ CB states -- shown in Fig.~\ref{fig:ordered_supercells_bands}(d) -- analogously to the calculation of the $\Gamma_{2'c}$ character. Doing so, we indeed find that the Ge$_{127}$C$_{1}$ CB edge eigenstate -- highlighted in Fig.~\ref{fig:ordered_supercells_bands}(d) via red coloring -- is constituted primarily of $\vert \textrm{L}_{1c}^{(0)} ( A_{1} ) \rangle$, having 70.3\% Ge L$_{1c} ( A_{1} )$ character. Finally, since the calculated Ge $\Gamma_{2'c}$ and L$_{1c} ( A_{1} )$ character of Figs.~\ref{fig:ordered_supercells_bands}(c) and ~\ref{fig:ordered_supercells_bands}(d) demonstrates the presence of hybridisation between $\Gamma_{2'c}$, L$_{1c} ( A_{1} )$ and higher energy CB states, we investigate the possibility of C-induced mixing with higher energy Ge host matrix CB states. In Ge$_{128}$ the next highest energy CB states above $\Gamma_{2'c}$ are the folded X$_{1c}$ CB edge states. As such, we construct $\vert \textrm{X}_{1c}^{(0)} ( A_{1} ) \rangle$ -- i.e.~a linear combination of Ge X$_{1c}$ states having purely $s$-like orbital character at the C lattice site -- and calculate the Ge X$_{1c} ( A_{1} )$ character of the Ge$_{127}$C$_{1}$ CB states. The results of this analysis are shown in Fig.~\ref{fig:ordered_supercells_bands}(e), where we note that the Ge$_{127}$C$_{1}$ CB edge eigenstate -- highlighted in Fig.~\ref{fig:ordered_supercells_bands}(e) via blue coloring -- acquires only 0.8\% Ge X$_{1c} ( A_{1} )$ character. We therefore conclude for the Ge$_{127}$C$_{1}$ supercell considered here (i) that C incorporation drives hybridisation between $A_{1}$-symmetric Ge host matrix states lying close in energy to the CB edge, (ii) that the Ge$_{127}$C$_{1}$ CB edge is derived primarily from a linear combination $\vert \textrm{L}_{1c}^{(0)} ( A_{1} ) \rangle$ of Ge L$_{1c}$ CB minimum states, and (iii) that the alloy band gap retains primarily indirect character.

% Ge(63)C(1) band structure

To investigate these trends as a function of $x$ we have repeated this analysis for an ordered, 64-atom Ge$_{63}$C$_{1}$ ($x = 1.56$\%) supercell, the calculated CB structure of which is shown in Fig.~\ref{fig:ordered_supercells_bands}(g). The CB structure of the corresponding C-free Ge$_{64}$ supercell is shown in Fig.~\ref{fig:ordered_supercells_bands}(f). Again, we note that the lowest energy CB states at $\textbf{K} = 0$ in Ge$_{64}$ are the folded L$_{1c}$ CB minimum states. Since this supercell has SC lattice vectors, the Brillouin zone boundary along (001) lies at $K_{z} = \frac{ \pi }{ A }$. Again, we note that C incorporation leads to a strong reduction of the band gap, by 93 meV compared to the fundamental band gap of Ge. Despite that the C composition in Ge$_{63}$C$_{1}$ is twice that in Ge$_{127}$C$_{1}$, we note that the calculated band gap reduction in both cases is approximately equal. This suggests strong composition-dependent bowing of the alloy band gap, the details of which depend precisely on the band mixing (hybridisation) present in a given alloy supercell. We note that the ordering, degeneracy and orbital character of the Ge$_{63}$C$_{1}$ $\textbf{K} = 0$ eigenstates are as described above for Ge$_{127}$C$_{1}$. Figures~\ref{fig:ordered_supercells_bands}(h),~\ref{fig:ordered_supercells_bands}(i) and~\ref{fig:ordered_supercells_bands}(j) show, respectively, the Ge $\Gamma_{2'c}$, L$_{1c} ( A_{1} )$ and X$_{1c} ( A_{1} )$ character of the Ge$_{63}$C$_{1}$ $\textbf{K} = 0$ CB states. Qualitatively, we note similar trends as for the Ge$_{127}$C$_{1}$ supercell. Firstly, C incorporation drives hybridisation between the $\Gamma_{2'c}$, L$_{1c}$ and X$_{1c}$ states of the Ge host matrix. Secondly, the alloy CB edge eigenstate retains primarily (77.9\%) indirect (Ge L$_{1c} ( A_{1} )$) character, and acquires only a small admixture (14.4\%) of direct (Ge $\Gamma_{2'c}$) character. As in Figs.~\ref{fig:ordered_supercells_bands}(c) --~\ref{fig:ordered_supercells_bands}(e), the $\Gamma_{2'c}$, L$_{1c} ( A_{1} )$ and X$_{1c} ( A_{1} )$ character of the Ge$_{63}$C$_{1}$ CB edge eigenstate in Figs.~\ref{fig:ordered_supercells_bands}(h) --~\ref{fig:ordered_supercells_bands}(j) is highlighted using green, red and blue coloring. Thirdly, C-induced hybridisation between Ge $\Gamma_{2'c}$ and X$_{1c}$ states is negligible, with the alloy CB edge state acquiring only minimal (2.2\%) Ge X$_{1c} ( A_{1} )$ character.

% Ge(53)C(1) band structure

We have so far considered the impact of C incorporation on the CB edge states only in the case where the L-point eigenstates of the corresponding C-free Ge supercell fold to $\textbf{K} = 0$. It is however pertinent to enquire as to whether the trends identified above -- and hence our conclusion that the Ge$_{1-x}$C$_{x}$ band gap retains primarily indirect character -- is a consequence of the choice of supercells employed in our analysis. Indeed, the mechanism driving the electronic structure evolution in response to C incorporation should be present in all alloy supercells, regardless of the specific choice of supercell. To address this issue we present also the results of equivalent analysis for an ordered, 54-atom Ge$_{53}$C$_{1}$ ($x = 1.85$\%) supercell, the calculated CB structure of which is shown in Fig.~\ref{fig:ordered_supercells_bands}(l). The CB structure of the corresponding C-free Ge$_{54}$ supercell is shown in Fig.~\ref{fig:ordered_supercells_bands}(k). For this $3 \times 3 \times 3$ FCC supercell the L points of the Brillouin zone of the underlying diamond lattice map directly to the L points of the supercell Brillouin zone. As such, the calculated CB minimum in Ge$_{54}$ lies at $\textbf{K} = ( \frac{ \pi }{ A }, \frac{ \pi }{ A }, \frac{ \pi }{ A } )$. Comparing Figs.~\ref{fig:ordered_supercells_bands}(k) and~\ref{fig:ordered_supercells_bands}(l) we note that, qualitatively, the impact of C incorporation on the CB edge again appears to illustrate the emergence of a direct band gap: the CB minimum relocates from $\textbf{K} = ( \frac{ \pi }{ A }, \frac{ \pi }{ A }, \frac{ \pi }{ A } )$ to $\textbf{K} = 0$ in response to C incorporation. Figure~\ref{fig:ordered_supercells_bands}(m) shows the calculated Ge $\Gamma_{2'c}$ character of the Ge$_{53}$C$_{1}$ $\textbf{K} = 0$ CB states. We note that the alloy CB edge again acquires only minority (22.0\%) Ge $\Gamma_{2'c}$ character. Despite the explicit choice of a Ge host matrix supercell in which the CB minimum is not folded to $\textbf{K} = 0$ (cf.~Fig.~\ref{fig:ordered_supercells_bands}(k)), giving way to a C-containing alloy supercell CB structure having its minimum at $\textbf{K} = 0$ (cf.~Fig.~\ref{fig:ordered_supercells_bands}(l)), careful analysis of the corresponding supercell CB edge eigenstate reveals primarily indirect character.

% Figure 3

\begin{figure*}[ht!]
	\includegraphics[width=1.00\textwidth]{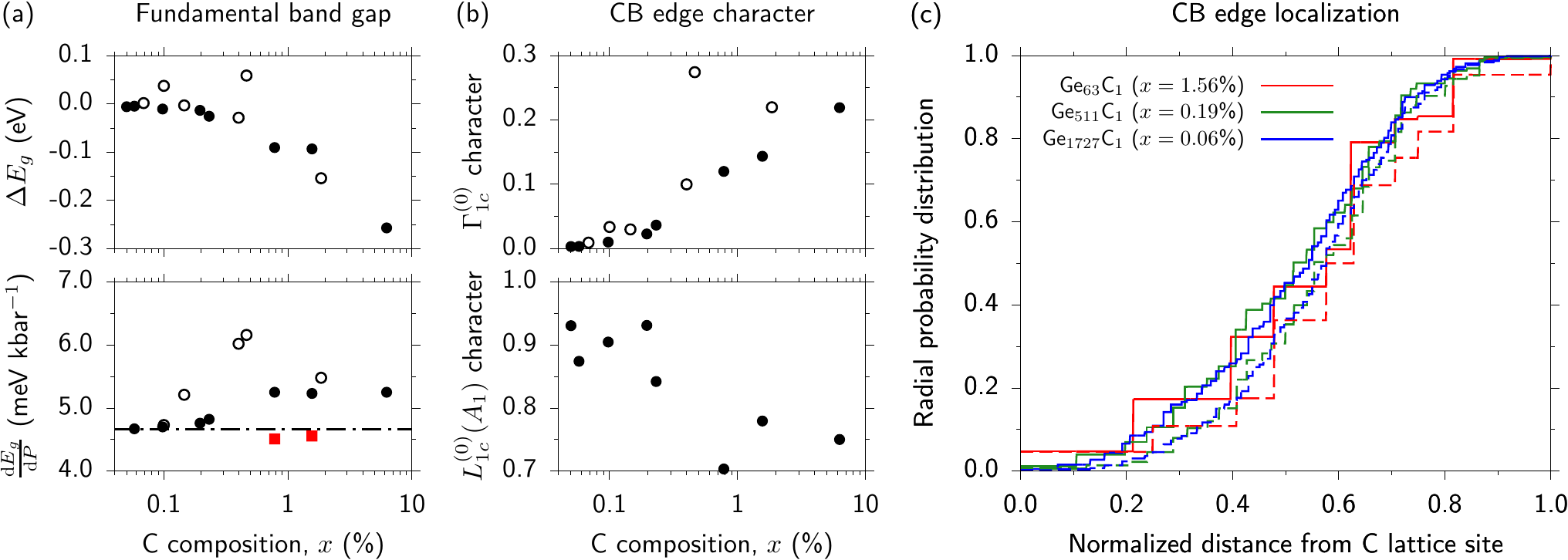}
	\caption{(a) Change in $\textbf{K} = 0$ band gap $\Delta E_{g}$ (upper panel) and band gap pressure coefficient $\frac{ \textrm{d} E_{g} }{ \textrm{d} P }$ (lower panel) as a function of C composition $x$ for a series of ordered Ge$_{N-1}$C$_{1}$ ($x = \frac{1}{N}$) alloy supercells containing $16 \leq N \leq 2000$ atoms. The dashed grey line in the lower panel denotes the pressure coefficient $\frac{ \textrm{d} E_{g} }{ \textrm{d} P } = 4.66$ meV kbar$^{-1}$ of the fundamental (indirect) band gap of Ge. The closed red squares in the lower panel show the hybrid functional DFT-calculated values of $\frac{ \textrm{d} E_{g} }{ \textrm{d} P }$ from Ref.~\onlinecite{Kirwan_SST_2019}. (b) Fractional Ge $\Gamma_{2'c}$ (upper panel) and L$_{1c} ( A_{1} )$ (lower panel) character of the alloy CB edge eigenstate for the same ordered Ge$_{N-1}$C$_{1}$ supercells as in (a). Closed (open) circles in (a) and (b) correspond to $n \times n \times n$ FCC or SC supercells having even (odd) values of $n$. (c) Radial probability distribution functions (RPDFs) for the CB edge eigenstate in ordered Ge$_{63}$C$_{1}$ (solid red line), Ge$_{511}$C$_{1}$ (solid green line) and Ge$_{1727}$C$_{1}$ (solid blue line) supercells. In each case the dashed line of the same color shows the calculated RPDF for the CB edge eigenstate $\vert \textrm{L}_{1c}^{(0)} ( A_{1} ) \rangle$ of the corresponding Ge$_{N}$ host matrix supercell.}
	\label{fig:ordered_supercells_summary}
\end{figure*}

In this 54-atom supercell hybridisation between the Ge $\Gamma_{2'c}$ and L$_{1c}$ states is blocked, since the latter do not fold back to $\textbf{K} = 0$. In Ge$_{54}$ the lowest energy CB states that fold to $\textbf{K} = 0$ originate from wave vectors $\textbf{k} = ( \frac{ 2 \pi }{ 3 a }, \frac{ 2 \pi }{ 3 a }, \frac{ 2 \pi }{ 3 a } )$, and equivalent points, located two-thirds of the way along the $\Lambda$ direction between the $\Gamma$ and L points in the primitive cell Brillouin zone of the underlying diamond lattice. These states, which we denote by $\vert \frac{2}{3} \Lambda_{1c}^{(0)} \rangle$, lie 33 meV above the $\Gamma_{2'c}$ zone-centre CB edge state in Fig.~\ref{fig:ordered_supercells_bands}(k). The next lowest energy set of Ge states which fold to $\textbf{K} = 0$ originate from wave vectors $\textbf{k} = ( 0, 0, \frac{ 4 \pi }{ 3 a } )$, and equivalent points, which lie two-thirds of the way along the $\Delta$ direction. The corresponding $\textbf{K} = 0$ states in Ge$_{54}$, which we denote by $\vert \frac{ 2 }{ 3 } \Delta_{1c}^{(0)} \rangle$, lie 88 meV above the $\Gamma_{2'c}$ zone-centre CB edge states. Figures~\ref{fig:ordered_supercells_bands}(n) and~\ref{fig:ordered_supercells_bands}(o) respectively show the corresponding calculated Ge $\frac{ 2 }{ 3 } \Lambda_{1c} ( A_{1} )$ and $\frac{ 2 }{ 3 } \Delta_{1c} ( A_{1} )$ character of the Ge$_{53}$C$_{1}$ $\textbf{K} = 0$ CB eigenstates. The Ge $\Gamma_{2'c}$, $\frac{ 2 }{ 3 } \Lambda_{1c} ( A_{1} )$ and $\frac{ 2 }{ 3 } \Delta_{1c} ( A_{1} )$ character of the CB edge eigenstate is highlighted in Figs.~\ref{fig:ordered_supercells_bands}(m) --~\ref{fig:ordered_supercells_bands}(o) using green, red and blue coloring, respectively. We calculate that the CB edge eigenstate has majority (70.2\%) Ge $\frac{2}{3} \Lambda_{1c} ( A_{1} )$ character, and acquires only minimal (0.8\%) Ge $\frac{2}{3} \Delta_{1c} ( A_{1} )$ character. We note the quantitative similarity to the Ge$_{127}$C$_{1}$ and Ge$_{63}$C$_{1}$ cases: despite that the L- and X-points do not fold to $\textbf{K} = 0$ in Ge$_{54}$, the Ge$_{53}$C$_{1}$ is nonetheless formed primarily of a linear combination of states originating from a point along the $\Lambda$ direction in the primitive cell Brillouin zone, which possesses purely $s$-like orbital character at the C lattice site. Overall, we therefore conclude that the CB edge eigenstate in ordered Ge$_{1-x}$C$_{x}$ alloys retains primarily indirect character, irrespective of the specific choice of supercell(s) employed in electronic structure calculations. As such, we consequently conclude that C incorporation does not drive the formation of a direct band gap in dilute Ge$_{1-x}$C$_{x}$. This conclusion is in direct agreement with the hybrid functional DFT calculations of Kirwan et al. \cite{Kirwan_SST_2019}

%%%%%%%%%%%%%%%%%%%%%%%%%%%%%%%%%%%%%%%%%%%%%%%%%%%%%%%%%%%%%%%%%%%%%%%%%%%%%%%%%%%%%%%%%%%%%%
%%%% Results: Trends vs.~C composition: band mixing and the ultra-dilute (impurity) limit %%%%
%%%%%%%%%%%%%%%%%%%%%%%%%%%%%%%%%%%%%%%%%%%%%%%%%%%%%%%%%%%%%%%%%%%%%%%%%%%%%%%%%%%%%%%%%%%%%%

\subsection{Trends vs.~C composition: band mixing and the ultra-dilute (impurity) limit}
\label{sec:results_impurity_limit}

% Summary of results of ordered supercells: Figs. 3(a) -- 3(d)

Having investigated in detail the impact of C incorporation in small ordered supercells containing $\leq 128$ atoms, we turn our attention now to the evolution of the alloy CB edge state and band gap as we approach the ultra-dilute limit of having an isolated substitutional C atom in a Ge matrix. The results of these calculations are summarised in Fig.~\ref{fig:ordered_supercells_summary}, which shows the calculated variation with $x$ of (a) the band gap reduction $\Delta E_{g}$ (upper panel) and band gap pressure coefficient $\frac{ \textrm{d} E_{g} }{ \textrm{d} P }$ (lower panel), and (b) the Ge $\Gamma_{2'c}$ character (upper panel) and Ge L$_{1c} ( A_{1} )$ (lower panel) character of the $\textbf{K} = 0$ CB edge eigenstate. Results are shown for ordered Ge$_{N-1}$C$_{1}$ supercells containing $16 \leq N \leq 2000$ atoms: the lowest (highest) C composition investigated is $x = 0.05$\% ($x = 6.25$\%) in an ordered $10 \times 10 \times 10$ FCC Ge$_{1999}$C$_{1}$ ($2 \times 2 \times 2$ FCC Ge$_{15}$C$_{1}$) supercell. Results for $n \times n \times n$ supercells having even (odd) values of $n$ are denoted in Figs.~\ref{fig:ordered_supercells_summary}(a) --~\ref{fig:ordered_supercells_summary}(b) by closed (open) circles. We recall that for odd $n$ supercells the L points of the primitive cell Brillouin zone associated with the underlying diamond lattice do not fold to $\textbf{K} = 0$, so that C-induced $\Gamma_{2'c}$-L$_{1c}$ hybridisation is blocked.

Examining Fig.~\ref{fig:ordered_supercells_summary}(a) we note that, in general, C incorporation drives a strong reduction of the $\textbf{K} = 0$ supercell band gap, with the magnitude of the band gap reduction $\Delta E_{g}$ growing strongly with increasing $x$ (decreasing $N$). The lowest band gap we calculate is $E_{g} = 0.599$ eV in a Ge$_{15}$C$_{1}$ supercell, which is reduced by 257 meV compared to the fundamental (indirect) band gap of Ge. For some large (ultra-dilute) supercells we calculate a supercell zone-centre ($\textbf{K} = 0$) band gap which exceeds the fundamental L$_{1c}$-$\Gamma_{25'v}$ band gap of the Ge host matrix. We emphasise that this is a result of the L$_{1c}$ states not folding to $\textbf{K} = 0$ in these supercells: the lowest energy CB states in these supercells originate from wave vectors $\textbf{k} = ( \frac{ 2 \pi }{ na }, \frac{ 2 \pi }{ na }, \frac{ 2 \pi }{ na } )$ located two-$n^{\scalebox{0.7}{\textrm{th}}}$s of the way along the $\Lambda$ direction in the primitive unit cell Brillouin zone, and hence lie higher in energy than the L$_{1c}$ CB minima. For sufficiently low $x$, C-induced hybridisation of $A_{1}$-symmetric Ge CB edge states is insufficient to push the fundamental band gap below that of Ge. Closed red squares in the bottom panel of Fig.~\ref{fig:ordered_supercells_summary}(a) show the values of $\frac{ \textrm{d} E_{g} }{ \textrm{d} P }$ calculated via hybrid functional DFT by Kirwan et al. \cite{Kirwan_SST_2019} Generally, we find that our TB-calculated pressure coefficients are $\approx 1$ meV kbar$^{-1}$ larger than those obtained from equivalent hybrid functional DFT calculations. This indicates a tendency of the TB calculations to overestimate the Ge $\Gamma_{2'c}$ character associated with the Ge$_{1-x}$C$_{x}$ alloy CB edge, emphasising the robustness of our conclusions regarding the indirect nature of the alloy band gap. Turning our attention to the lower part of Fig.~\ref{fig:ordered_supercells_summary}(b), we again note that the precise calculated value of $\frac{ \textrm{d} E_{g} }{ \textrm{d} P }$ depends on the specific details of band hybridisation -- as determined by zone folding -- in a given alloy supercell. For all supercells the calculated values of $\frac{ \textrm{d} E_{g} }{ \textrm{d} P }$ remain significantly closer to the value 4.66 meV kbar$^{-1}$ associated with the L$_{1c}$-$\Gamma_{25'v}$ indirect band gap of Ge, than to the value 13.33 meV kbar$^{-1}$ associated with the direct $\Gamma_{2'c}$-$\Gamma_{25'v}$ band gap. This emphasises that the Ge$_{1-x}$C$_{x}$ band gap retains primarily indirect character, even for C compositions as high as $x = 6.25$\%.

The indirect nature of the Ge$_{1-x}$C$_{x}$ band gap is further emphasised by analysing the evolution of the calculated Ge $\Gamma_{2'c}$ and L$_{1c} ( A_{1} )$ character of the Ge$_{N-1}$C$_{1}$ CB edge eigenstates, shown respectively in the upper and lower panels of Fig.~\ref{fig:ordered_supercells_summary}(b). Note the absence of open circles in the lower panel of Fig.~\ref{fig:ordered_supercells_summary}(b): since the L-point eigenstates in an odd $n$ supercell do not fold to $\textbf{K} = 0$, the Ge L$_{1c}$ eigenstates can not contribute to the $\textbf{K} = 0$ supercell eigenstates. As $x$ decreases ($N$ increases) towards the ultra-dilute limit, we note that the CB edge Ge $\Gamma_{2'c}$ character tends towards zero. Similarly, we note that the CB edge Ge L$_{1c} ( A_{1} )$ character tends to increase with decreasing $x$. While the CB edge Ge $\Gamma_{2'c}$ character reaches a value as low as 0.3\% in the largest (Ge$_{1999}$C$_{1}$) supercell considered, we note that the corresponding Ge L$_{1c} ( A_{1} )$ character attains a value 93.1\% in the same supercell. This reflects that, for the case of an isolated substitutional C atom, the CB edge state consists primarily of a linear combination of Ge L$_{1c}$ states having $A_{1}$ symmetry at the C lattice site, and also possesses a small admixture of other (non-$\Gamma_{2'c}$) $A_{1}$-symmetric linear combinations of Ge states. The largest calculated alloy CB edge Ge $\Gamma_{2'c}$ character in an even $n$ supercell is 21.9\% in Ge$_{15}$C$_{1}$ ($x = 6.25$\%), suggesting that Ge$_{1-x}$C$_{x}$ acquires minimal direct-gap character even as the C composition is increased beyond the dilute regime. We note higher Ge $\Gamma_{2'c}$ character in the 216-atom Ge$_{215}$C$_{1}$ ($x = 0.46$\%) supercell, but recall from Fig.~\ref{fig:ordered_supercells_summary}(a) that the lowest energy CB state at $\textbf{K} = 0$ in this supercell lies above the Ge L$_{1c}$ CB edge, with the resultant reduced separation in energy between the folded Ge$_{216}$ CB edge and the $\Gamma_{2'c}$ state driving strong hybridisation in a 216-atom alloy supercell.

For small ($N \leq 128$) supercells we note that our calculated values of $\Delta E_{g}$ and $\frac{ \textrm{d} E_{g} }{ \textrm{d} P }$ are in good quantitative agreement with the hybrid DFT calculations of Ref.~\onlinecite{Kirwan_SST_2019}. Our calculated CB structure for an ordered Ge$_{127}$C$_{1}$ supercell is also in good overall agreement with the calculations of Kirwan et al.~\cite{Kirwan_SST_2019}, as well as those of Stephenson et al.~\cite{Stephenson_JEM_2016}. At higher C compositions our calculations disagree qualitatively with the hybrid DFT calculations of Ref.~\onlinecite{Stephenson_JEM_2016}, which suggest that the alloy band gap has closed even in a Ge$_{53}$C$_{1}$ ($x = 1.85$\%) supercell. We note however that the Ge$_{53}$C$_{1}$ supercell band structure presented in Ref.~\onlinecite{Stephenson_JEM_2016} contains several unusual features, including a large energy splitting of the VB edge states. This VB edge splitting, which should vanish in an ordered alloy supercell, suggests improper supercell relaxation, castng doubt on the suggestion that the band gap closes in the C composition range between 1.56\% (in Ge$_{63}$C$_{1}$) and 1.85\% (in Ge$_{53}$C$_{1}$).

%%%%%%%%%%%%%%%%%%%%%%%%%%%%%%%%%%%%%%%%%%%%%%%%%%%%%%%%%%%%%%%%%%%%%%%%%%%%%%%%%%%%%%%%%%%%
%%%% Results: Evolution of conduction band edge eigenstates in ordered alloy supercells %%%%
%%%%%%%%%%%%%%%%%%%%%%%%%%%%%%%%%%%%%%%%%%%%%%%%%%%%%%%%%%%%%%%%%%%%%%%%%%%%%%%%%%%%%%%%%%%%

\subsection{Evolution of conduction band edge eigenstates in ordered alloy supercells}
\label{sec:results_localisation}

Having considered in detail the character of the CB edge eigenstates, we finally consider the potential for carrier localisation in response to C incorporation. The solid red, green and blue lines in Fig.~\ref{fig:ordered_supercells_summary}(c) respectively show the calculated cumulative radial probability distribution function (RPDF) associated with the CB edge eigenstates in Ge$_{63}$C$_{1}$ ($x = 1.56$\%), Ge$_{511}$C$_{1}$ ($x = 0.19$\%) and Ge$_{1727}$C$_{1}$ ($x = 0.06$\%) -- i.e.~$2 \times 2 \times 2$, $4 \times 4 \times 4$ and $6 \times 6 \times 6$ SC -- supercells. Here, the cumulative RPDF is calculated for a given supercell eigenstate by selecting the C lattice site as the origin of coordinates and then, at a given distance from this origin, adding the total probability density residing on atoms located at that distance from the C lattice site. For a given supercell eigenstate the rate at which the cumulative RPDF approaches a value of unity gives an indication of the degree of localisation of the state. As such, for an eigenstate strongly localised about the C lattice site, we would expect the calculated cumulative RPDF in Fig.~\ref{fig:ordered_supercells_summary}(c) to rapidly approach a value of unity with increasing distance. To facilitate comparison of RPDFs calculated for eigenstates of supercells having different size, we plot the RPDF as a function of distance normalised to the maximum interatomic distance in each supercell -- i.e.~normalised by the dimensions of a given supercell. For a highly localised alloy CB edge state, we would then expect the cumulative RPDF to approach unity more rapidly with increasing supercell size in Fig.~\ref{fig:ordered_supercells_summary}(c).

To quantify the presence of any localisation generated by C incorporation, we show also in Fig.~\ref{fig:ordered_supercells_summary}(c) the cumulative RPDFs associated with the constructed host matrix CB edge eigenstate $\vert \textrm{L}_{1c}^{(0)} ( A_{1} ) \rangle$ for the Ge$_{64}$ (dashed red line), Ge$_{512}$ (dashed green line) and Ge$_{1728}$ (dashed blue line) supercells. We note that these $\vert \textrm{L}_{1c}^{(0)} ( A_{1} ) \rangle$ host matrix CB edge states extend through the full supercell: in a given $N$-atom Ge$_{N}$ supercell the associated cumulative RPDF therefore increases smoothly in magnitude with increasing distance from the lattice site on which the eigenstate has been constructed to have purely $s$-like orbital character. Considering firstly the cumulative RPDF associated with the Ge$_{64}$ CB edge eigenstate, we calculate that 10\% of the probability density associated with $\vert \textrm{L}_{1c}^{(0)} ( A_{1} ) \rangle$ resides on the origin and its four nearest-neighbour lattice sites. Substituting the Ge atom at the origin by C, this increases to 18\% for the corresponding Ge$_{63}$C$_{1}$ ($x = 1.56$\%) supercell. Based solely on analysis of small supercell eigenstates, it then appears that C incorporation generates appreciable electron localisation at the CB edge. However, this trend is not borne out in the larger supercells: in both Ge$_{511}$C$_{1}$ and Ge$_{1727}$C$_{1}$ we calculate that the associated increase in localisation compared to the $\vert \textrm{L}_{1c} ( A_{1} ) \rangle$ host matrix CB edge state is $< 5$\%.

To quantify the overall change in localisation in response to C incorporation, we have also calculated the inverse participation ratio (IPR) associated with the CB edge eigenstates in all supercells studied. \cite{Thouless_PR_1974,Tanner_RSCA_2016} The IPR associated with an eigenstate in an $N$-atom supercell attains a minimum value of $\frac{1}{N}$ for a fully delocalised eigenstate having equal probability density at each lattice site, and a maximum value of 1 for a fully localised eigenstate for which the probability density resides entirely at a single lattice site. The calculated IPR for the Ge$_{N-1}$C$_{1}$ CB edge eigenstate tends towards an average value $\approx \frac{2}{N}$ in the ultra-dilute limit, a minimal increase from the value $\frac{1.62}{N}$ calculated for the corresponding $\vert \textrm{L}_{1c}^{(0)} ( A_{1} ) \rangle$ host matrix eigenstate. Similar analysis for higher energy alloy CB states reveals qualitatively similar results: incorporation of an \textit{isolated} substitutional C atom in Ge does not lead to any significant electron localisation about the C atom. (As we will describe in Sec.~\ref{sec:results_disordered} below, this is no longer the case in the presence of C clustering, which can result in significant localisation of electrons occupying states lying energetically within the Ge band gap.)

The strong CB edge localisation observed in calculations for small ordered Ge$_{N-1}$C$_{1}$ supercells \cite{Kirwan_SST_2019} therefore gives way to a delocalised alloy CB edge in the ultra-dilute (large $N$) limit. This is in stark contrast to equivalent analysis of dilute nitride GaN$_{x}$(As,P)$_{1-x}$ alloys, \cite{Lindsay_PE_2004,Harris_JPCM_2008} to which Ge$_{1-x}$C$_{x}$ has recently been compared, where substitutional N incorporation generates N-related impurity states which are found to be highly localised about the N lattice site in the ultra-dilute limit. Based on the calculated high Ge L$_{1c} ( A_{1} )$ character of the Ge$_{1-x}$C$_{x}$ CB edge eigenstates (cf.~Fig.~\ref{fig:ordered_supercells_bands}(d) and~\ref{fig:ordered_supercells_bands}(f)), this lack of strong localisation is not surprising: the alloy CB edge is formed primarily from a linear combination of a small number of delocalised Ge eigenstates, and is hence itself delocalised. As such, we conclude that dilute C incorporation in Ge does \textit{not} introduce significant electron localisation about substitutional C atoms, which are spaced widely apart in ordered alloy supercells.

% At intermediate distances from the C lattice site we note that the cumulative RPDF associated with the alloy CB edge slightly exceeds that associated with $\vert \textrm{L}_{1c}^{(0)} ( A_{1} ) \rangle$. Bearing in mind the use of Born-von Karman boundary conditions in supercell calculations, we note that this results suggests that C-induced hybridisation of Ge host matrix states leads to a small -- but qualitatively unimportant -- increase in localisation of electrons between substitutional C atoms. It is likely that this intermediate distance behaviour is a consequence of the ordered supercells for which these calculations were performed and, as such, that preferential electron localisation away from substitutional C atoms would not be present in realistic, disordered Ge$_{1-x}$C$_{x}$ alloys.

% Conclusions based on ordered supercell calculations

On the basis of our analysis of the electronic structure of ordered Ge$_{1-x}$C$_{x}$ alloy supercells we conclude overall that -- despite the large mismatch in size and chemical properties between Ge and C -- from the perspective of the electronic structure an \textit{isolated} C atom does not act as an isovalent impurity when incorporated substitutionally in Ge. Our analysis explicitly rules out the interpretation of the Ge$_{1-x}$C$_{x}$ CB structure in terms of C-related localised impurity states and the BAC model. Moreover, our analysis therefore suggests --  in agreement with the conclusions of Kirwan et al.~\cite{Kirwan_SST_2019} -- that substitutional C incorporation in Ge does \textit{not} drive the formation of a direct band gap in dilute Ge$_{1-x}$C$_{x}$ alloys.

%%%%%%%%%%%%%%%%%%%%%%%%%%%%%%%%%%%%%%%%%%%%%%%%%%%%%%%%%%%%%%%%%%%%%%%%%%%%%%%%%%%%%%%%%%%%%%%%%%%%
%%%% Results: Impact of C incorporation on conduction band edge states in disordered GeC alloys %%%%
%%%%%%%%%%%%%%%%%%%%%%%%%%%%%%%%%%%%%%%%%%%%%%%%%%%%%%%%%%%%%%%%%%%%%%%%%%%%%%%%%%%%%%%%%%%%%%%%%%%%

\subsection{Impact of C incorporation on conduction band edge states in disordered Ge$_{1-x}$C$_{x}$ alloys}
\label{sec:results_disordered}

% \begin{itemize}
% 	\item Try to make connection to spectroscopic measurements (Wistey et al., 2016 J.~Appl.~Phys.)
% \end{itemize}

% Figure 4

\begin{figure*}[ht!]
	\includegraphics[width=1.00\textwidth]{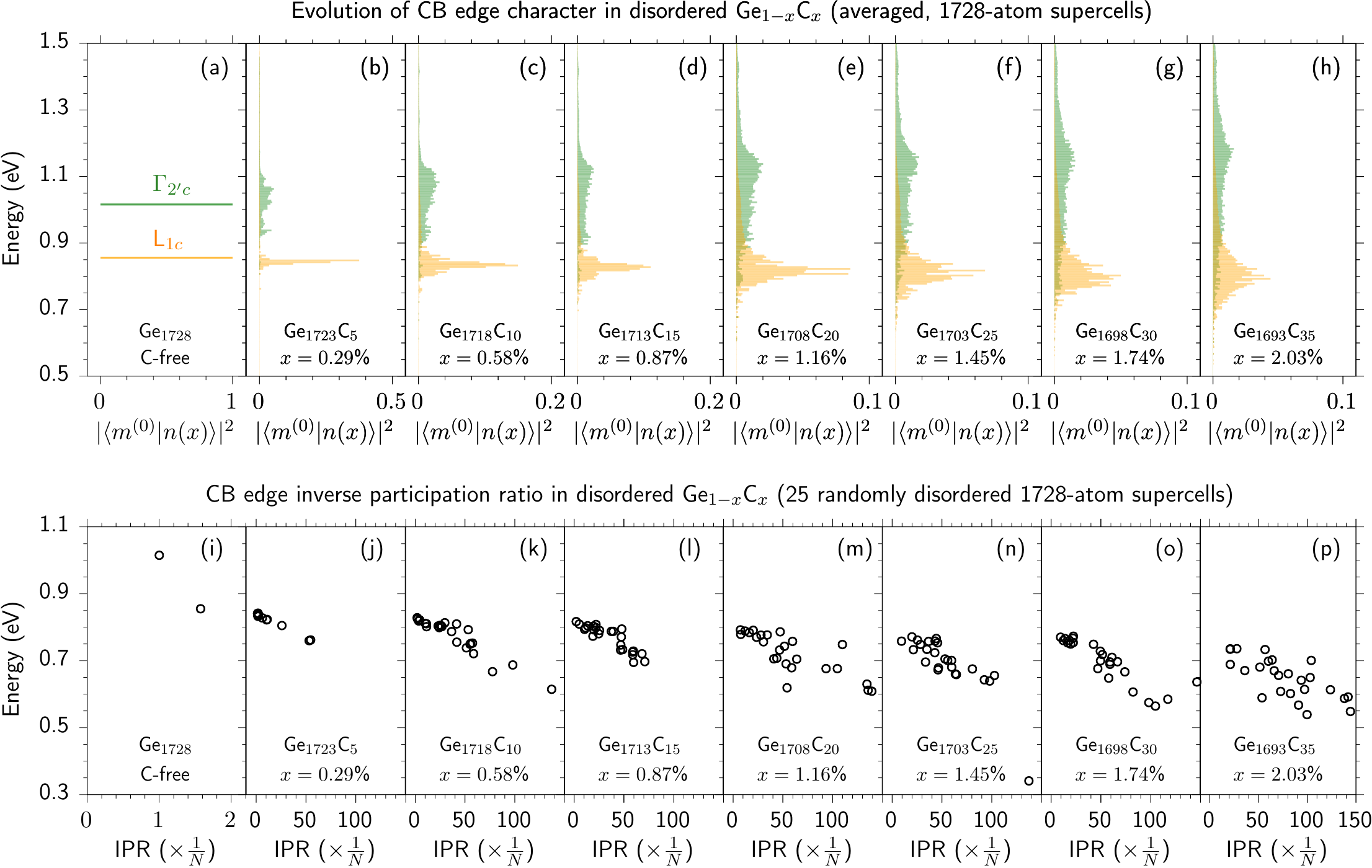}
	\caption{Top row: Evolution of the character of the alloy CB edge states in 1728-atom disordered Ge$_{1-x}$C$_{x}$ supercells. (a) Fractional Ge $\Gamma_{2'c}$ (green) and L$_{1c}$ (orange) character spectra for a C-free Ge$_{1728}$ ($6 \times 6 \times 6$ SC) supercell, calculated respectively by projecting the Ge$_{1728}$ CB edge eigenstate $\vert m^{(0)} \rangle = \vert \Gamma_{2'c}^{(0)} \rangle$ or $\vert \textrm{L}_{1c}^{(0)} \rangle$ onto the full set of supercell zone-centre ($\textbf{K} = 0$) eigenstates. (b) -- (h) respectively show the fractional Ge $\Gamma_{2'c}$ and L$_{1c}$ character spectra for a series of disordered, 1728-atom Ge$_{1728-M}$C$_{M}$ supercells having $M = 5$ -- 35, in steps of 5 C atoms, corresponding respectively to C compositions $x = 0.29$\%, 0.58\%, 0.87\%, 1.16\%, 1.45\%, 1.74\% and 2.03\%. Note the difference in scales on the abscissae of panels (a), (b), (c) -- (d), and (e) -- (h). Bottom row: Calculated energy and IPR of the CB edge eigenstate in the 25 distinct, randomly disordered 1728-atom Ge$_{1728-M}$C$_{M}$ supercells for which averaged data are presented in the top row. (i) shows the calculated energies and IPRs of the $\vert \Gamma_{2'c}^{(0)} \rangle$ and $\vert \textrm{L}_{1c}^{(0)} ( A_{1} ) \rangle$ states in C-free Ge$_{1728}$. (j) -- (p) respectively show the alloy CB edge energies and IPRs calculated for supercells having C compositions $x = 0.29$\%, 0.58\%, 0.87\%, 1.16\%, 1.45\%, 1.74\% and 2.03\%. Note the difference in scales on the abscissae of panels (i) and (j) -- (p).}
	\label{fig:disordered_supercells_character}
\end{figure*}

% Choice of supercells

Having quantified the impact of C incorporation on the electronic structure of ordered Ge$_{1-x}$C$_{x}$ alloy supercells, we turn our attention now to more realistic, disordered alloy supercells. To analyse the electronic structure in the presence of alloy disorder, two key requirements guide our choice of supercells. Firstly, we must choose sufficiently large supercells so that incorporation of dilute C compositions $x \sim 1$\% correspond to substitution of multiple C atoms. This allows for the formation of disordered local neighbour environments in the supercell, and hence allows alloy disorder effects to be explicitly included in our electronic structure calculations. Secondly, since we have identified that C incorporation drives hybridisation between Ge host matrix states lying close in energy to the CB edge, it is important that we choose supercells which accurately reflect this band mixing as it occurs in a real Ge$_{1-x}$C$_{x}$ alloy. In supercell calculations, more states fold back to $\textbf{K} = 0$ close in energy to the CB edge as the supercell size increases, providing a more accurate representation of C-induced band hybridisation. Based on the analysis of Sec.~\ref{sec:results_impurity_limit} -- showing that the electronic properties associated with an isolated C impurity have converged for $N \approx 2000$ atoms in the supercell -- we therefore use 1728-atom ($6 \times 6 \times 6$ SC) Ge$_{1728-M}$C$_{M}$ ($x = \frac{ M }{ 1728 }$) supercells, and perform calculations for supercells containing up to $M = 35$ substitutional C atoms ($x \approx 2$\%). To obtain a reliable description of the evolution of the electronic structure with $x$, we perform a high-throughput analysis: at each C composition we construct, relax and calculate the electronic structure of 25 distinct disordered supercells, where Ge atoms are substituted by C at randomly selected lattice sites. Results for a given C composition are then obtained via configurational averaging -- i.e.~by averaging over the results of calculations for these 25 distinct disordered supercells.

The configurational averaging approach is favourable here compared to the use of a single, ultra-large supercell at each distinct C composition, due to the localised nature of the alloy CB states in \textit{disordered} Ge$_{1-x}$C$_{x}$ alloys (described below). In conventional semiconductor alloys, which display minimal carrier localisation, eigenstate formation arises from the (generally weak) hybridisation of delocalised eigenstates, therefore requiring ultra-large supercells to quantitatively describe the resultant alloy properties by directly encapsulating the associated length scales. \cite{Zhang_PRB_2011} Conversely, highly-mismatched alloys are characterised by effects pertaining to carrier localisation and short-range alloy disorder, which act on significantly reduced length scales. \cite{Kent_SST_2002,Reilly_SST_2009,Broderick_nitride_chapter_2017,Broderick_bismide_chapter_2017} Supercells containing $N \sim 10^{3}$ atoms are therefore sufficient to describe the properties of such materials, with configurational averaging providing an efficient means to explore the impact of the formation of distinct short-range (near-neighbour) environments on the electronic structure. \cite{Kent_PRB_2001,Kent_SST_2002,Usman_PRB_2013,Broderick_PRB_2014}

% Evolution of alloy conduction band edge character

As in Secs.~\ref{sec:results_ordered} --~\ref{sec:results_localisation} above, our primary concern is to quantify the evolution of the character of the alloy CB states with increasing $x$. To do this, we use the $\Gamma_{2'c}$ and folded L$_{1c}$ states $\vert \Gamma_{2'c}^{(0)} \rangle$ and $\vert \textrm{L}_{1c}^{(0)} \rangle$ of the Ge$_{1728}$ host matrix supercell to calculate the fractional Ge $\Gamma_{2'c}$ and L$_{1c}$ character spectra of the Ge$_{1728-M}$C$_{M}$ $\textbf{K} = 0$ CB eigenstates. To average the calculated spectra at fixed $x$ we sort the calculated Ge $\Gamma_{2'c}$ and L$_{1c}$ character into energy intervals of width 5 meV: for each distinct disordered supercell a given energy interval is populated by the total Ge $\Gamma_{2'c}$ or L$_{1c}$ character of alloy CB states lying in the range of energy spanned by the interval, and the resulting totals for all 25 supercells are then averaged to obtain the average Ge $\Gamma_{2'c}$ or L$_{1c}$ character in that energy range. The results of this analysis are summarised in Figs.~\ref{fig:disordered_supercells_character}(a) --~\ref{fig:disordered_supercells_character}(h), which show the calculated evolution with $x$ of the averaged fractional Ge $\Gamma_{2'c}$ (green) and L$_{1c}$ (orange) character for the disordered 1728-atom Ge$_{1-x}$C$_{x}$ supercells described above. The C compositions in Fig.~\ref{fig:disordered_supercells_character} begin at $x = 0$ for the C-free Ge$_{1728}$ host matrix supercell in Fig.~\ref{fig:disordered_supercells_character}(a), and increase in steps of $M = 5$ C atoms in subsequent panels up to a maximum C composition $x = 2.03$\% ($M = 35$) in fig.~\ref{fig:disordered_supercells_character}(h). Note the different scales on the abscissae of Figs.~\ref{fig:disordered_supercells_character}(a),~\ref{fig:disordered_supercells_character}(b), and~\ref{fig:disordered_supercells_character}(c) --~\ref{fig:disordered_supercells_character}(h).

Beginning in Fig~\ref{fig:disordered_supercells_character}(a) with the reference Ge$_{1728}$ supercell, the lowest energy $\textbf{K} = 0$ CB states are the fourfold degenerate folded L$_{1c}$ states which possess 100\% Ge L$_{1c}$ character, while the next lowest energy CB state is the $\Gamma_{2'c}$ state which possesses 100\% Ge $\Gamma_{2'c}$ character. Next, Fig.~\ref{fig:disordered_supercells_character}(b) shows the corresponding spectra calculated and averaged for Ge$_{1723}$C$_{5}$ ($x = 0.29$\%) supercells. We firstly observe, as in our ordered supercell calculations (cf.~Sec.~\ref{sec:results_ordered}), that C incorporation generates a downward shift in energy of CB states possessing Ge L$_{1c}$ character, reflecting the introduction of primarily Ge L$_{1c}$-derived states lying energetically within the Ge band gap. We also observe broadening of the energy range within which the Ge L$_{1c}$ character resides. This strong energetic broadening of the CB edge Bloch character is a consequence of C-related alloy disorder. In ordered alloy supercells containing only a single substitutional C impurity, the Born-von Karman boundary conditions generate an ordered alloy in which the C atoms are arranged on a regular grid -- determined by the supercell lattice vectors -- and the underlying cubic symmetry of the diamond lattice is preserved. The calculated Ge character for ordered supercells then resides only on discrete alloy states (cf.~Fig.~\ref{fig:ordered_supercells_bands}), with the associated spectral features having zero width in energy. Substitution of multiple C atoms at randomly selected lattice sites to form a disordered alloy supercell results in short-range structural disorder, breaking the underlying cubic symmetry. This reduction in symmetry, associated with the presense of a wide range of C local neighbour environments, leads in general to the lifting of degeneracies and, in the case of highly-mismatched alloys such as Ge$_{1-x}$C$_{x}$, to strong dependence of the calculated electronic properties, including the CB edge energy, on the precise short-range disorder present in a given alloy supercell. The Ge $\Gamma_{2'c}$ and L$_{1c}$ character is then broadened and spread over a continuous energy range in a real Ge$_{1-x}$C$_{x}$ alloy, with the degree of energetic broadening reflecting the sensitivity of the electronic structure to short-range structural disorder.

Turning our attention to the averaged Ge $\Gamma_{2'c}$ character in Fig.~\ref{fig:disordered_supercells_character}(b), we observe extremely strong energetic broadening. While the Ge L$_{1c}$-related feature remains sharply defined in energy, we find that alloy disorder gives rise to strong intrinsic inhomogeneous broadening of the Ge $\Gamma_{2'c}$ character. At this C composition $x = 0.29$\% we find that the maximum Ge $\Gamma_{2'c}$ character residing within a single 5 meV energy interval is 5.4\%. We note that this is in contrast to the results shown in Figs.~\ref{fig:ordered_supercells_bands}(c),~\ref{fig:ordered_supercells_bands}(h) and~\ref{fig:ordered_supercells_bands}(m) for small ordered supercells containing a single C atom, where the Ge $\Gamma_{2'c}$ character resides on only a small number of alloy CB levels. Our calculations therefore suggest that C-related alloy disorder leads to strong inhomogeneous energetic broadening of the Ge $\Gamma_{2'c}$ character. Similarly to the ordered supercell calculations of Sec.~\ref{sec:results_ordered}, we find that states lying in the same energy interval generally possess an admixture of both Ge $\Gamma_{2'c}$ and L$_{1c}$ character -- i.e.~they are hybridised states formed of a linear combination of Ge host matrix states. The $\Gamma_{2'c}$ character is found predominantly at higher energies, with only a limited amount of $\Gamma_{2'c}$ character mixing into the lowest energy CB states. By comparison, we find that the alloy supercell VB edge states (not shown) retain strong Ge $\Gamma_{25'v}$ character and display comparatively minor energetic broadening, reflecting that C incorporation has minimal impact on the VB structure.

% We note that our calculations here are in good qualitative agreement with photo-modulated reflectance (PR) spectra measured for a Ge$_{0.998}$C$_{0.002}$ ($x = 0.2$\%) sample. \cite{Stephenson_JAP_2016} Firstly, the PR feature associated with the fundamental band gap in Ge$_{0.998}$C$_{0.002}$ lies slightly lower in energy than, but is broadened in energy compared to, the PR feature associated with the fundamental (indirect) band gap of a Ge reference sample. This feature corresponds in our calculations to transitions between alloy CB states possessing Ge L$_{1c}$ character and comparatively unperturbed VB edge states, reflecting the indirect nature of the fundamental alloy band gap. Secondly, an additional PR feature emerges in the Ge$_{0.998}$C$_{0.002}$ PR spectrum, in an energy range $\approx 0.1$ eV above that associated with the fundamental band gap. This feature, which is both broader in energy and weaker than that associated with the fundamental band gap, corresponds in our calculations to transitions involving alloy CB states possessing Ge $\Gamma_{2'c}$ character.

As $x$ is further increased, we note the continuation of this general trend. Firstly, the feature associated with the Ge L$_{1c}$ character is both shifted downwards and increasingly broadened in energy with increasing $x$. Secondly, we observe extremely strong inhomogeneous broadening of the Ge $\Gamma_{2'c}$ character, which becomes spread over a broad range of energies at and above the CB edge in a given alloy supercell. Thirdly, we note that states lying close in energy to the alloy CB edge acquire minimal Ge $\Gamma_{2'c}$ character and remain primarily Ge L$_{1c}$-derived. This describes that the fundamental alloy band gap remains primarily indirect in character, similar to that in Ge, but that spectral features associated with this band gap undergo strong inhomogeneous broadening due to the presence of alloy disorder. Again, we note that this behaviour is markedly different to that expected on the basis of the BAC model, whereby a direct band gap would be expected to emerge via the transfer of increasing Ge $\Gamma_{2'c}$ character to the alloy CB edge with increasing C composition $x$. In general, the results of our disordered alloy analysis support the conclusions of Kirwan et al., \cite{Kirwan_SST_2019} and of our ordered supercell analysis in Sec.~\ref{sec:results_impurity_limit} above: dilute C incorporation in Ge does \textit{not} give rise to a direct band gap. Given the small $\Gamma_{2'c}$ character associated with individual Ge$_{1-x}$C$_{x}$ CB states, we do not expect strong optical matrix elements and direct-gap optical recombination between these states and the VB edge.

Finally, the calculated strong inhomogeneous broadening of the Ge $\Gamma_{2'c}$ and L$_{1c}$ character suggests the formation of a distribution of C-related localised states close in energy to the CB edge in Ge$_{1-x}$C$_{x}$ alloys. Given that an isolated C impurity generates minimal localisation at the CB edge (cf.~Sec.~\ref{sec:results_localisation}), this suggests that C-related alloy disorder -- i.e.~the formation of nearest-neighbour C-C pairs and even near-neighbour C pairs, as well as larger clusters of neighbouring C atoms -- can generate significant electron localisation. To ascertain the extent to which this is the case, at each C composition $x$ we have computed the IPR associated with the lowest energy CB eigenstate in each of the 25 randomly disordered Ge$_{1728-M}$C$_{M}$ supercells for which averaged data are presented in Figs.~\ref{fig:disordered_supercells_character}(a) --~\ref{fig:disordered_supercells_character}(h). The calculated energies and IPRs of these states are shown in Figs.~\ref{fig:disordered_supercells_character}(i) --~\ref{fig:disordered_supercells_character}(p). Recalling that a larger IPR reflects stronger localisation, we firstly note a general trend in the calculated IPRs for distinct supercells at fixed $x$: the IPR associated with a given eigenstate tends to increase strongly with decreasing eigenstate energy. This reflects the emergence of tightly-bound C-related cluster states lying deep within the Ge host matrix band gap in energy -- related predominantly here to the presence of C atoms in close proximity to one another -- similar to the distribution of N-related cluster states in dilute nitride GaN$_{x}$(As,P)$_{1-x}$. \cite{Kent_PRB_2001,Lindsay_PRL_2004} Secondly, we note that the localisation of the lowest energy CB state in Ge$_{1-x}$C$_{x}$ tends on average to increase with increasing $x$, reflecting the formation of more localised states in response to the closer proximity of substitutional C atoms at higher C compositions. Thirdly, at fixed $x$ we note the spread in energy of the lowest energy alloy CB state in distinct randomly disordered supercells, which tends to increase with increasing $x$ and is $\gtrsim 0.2$ eV for the highest C composition considered. This accounts for the strong inhomogeneous broadening of the CB states observed in Figs.~\ref{fig:disordered_supercells_character}(a) --~\ref{fig:disordered_supercells_character}(h), and confirms that the strong sensitivity of the electronic structure to short-range alloy disorder is a consequence of electron localisation about clusters of substitutional C atoms. As an extreme example of this localisation we note that the CB edge eigenstate in one of the Ge$_{1703}$C$_{25}$ ($x = 1.45$\%) supercells considered has both a markedly low energy of 0.341 eV and a high IPR of $\frac{137}{N}$ (visible in the bottom right-hand corner of Fig.~\ref{fig:disordered_supercells_character}(n)). Further inspection reveals that this highly localised CB edge eigenstate is associated with the presence of a C-Ge-C-Ge-C nearest-neighbour chain in the supercell, about which the calculated eigenstate is strongly localised (with its probability density distributed over only $\left( \frac{137}{N} \right)^{-1} \approx 13$ atoms in total).

%%%%%%%%%%%%%%%%%%%%%%%%%%%%%%%%%%%%%%%%%%%%%%
%%%% Implications for device applications %%%%
%%%%%%%%%%%%%%%%%%%%%%%%%%%%%%%%%%%%%%%%%%%%%%

\section{Implications for device applications}
\label{sec:implications}

Previous analysis has suggested the emergence of a direct band gap in dilute Ge$_{1-x}$C$_{x}$ alloys, and that the resulting alloy band structure should produce performance in Ge$_{1-x}$C$_{x}$-based semiconductor lasers and modulators which is comparable to that in conventional direct-gap III-V semiconductor materials. \cite{Stephenson_JAP_2016} Conversely, the results of our detailed analysis of electronic structure evolution in Ge$_{1-x}$C$_{x}$ alloys have negative implications from the perspective of potential device applications.

Firstly, our calculations for ordered alloy supercells suggest that the large differences in size and electronegativity between C and Ge drive hybridisation between Ge host matrix states lying close in energy to the CB edge. While this C-induced band mixing results in the transfer of some direct (Ge $\Gamma_{2'c}$) character to the alloy CB edge, this direct-gap character in general remains minimal. Indeed, we confirmed that dilute Ge$_{1-x}$C$_{x}$ admits a ``quasi-direct'' band gap: while supercell electronic structure calculations show the CB minimum to lie at $\textbf{K} = 0$, the alloy CB edge, and hence band gap, retains primarily indirect character.

Secondly, our analysis of large disordered alloy supercells also demonstrates strong sensitivity of the Ge$_{1-x}$C$_{x}$ electronic properties to the presence of short-range alloy disorder, behaviour typical of a highly-mismatched semiconductor alloy. In the presence of alloy disorder our calculations describe a breakdown of the CB edge Bloch character, leading to a broad distribution of C-related localised states lying below the Ge CB edge in energy. Individually, these states possess only minimal direct (Ge $\Gamma_{2'c}$) character, and so will not support appreciable direct-gap optical recombination. Further analysis would be required to estimate the magnitude of the optical recombination rate associated with these localised states in a realistic, disordered Ge$_{1-x}$C$_{x}$ alloy. Such analysis is beyond the scope of this work, but, based on our calculated $\Gamma_{2'c}$ character spectra (cf.~Fig.~\ref{fig:disordered_supercells_character}), we would expect a considerably lower radiative recombination rate in dilute Ge$_{1-x}$C$_{x}$ than in a conventional direct-gap semiconductor.

From the perspective of carrier transport, experimental measurements for closely-related dilute Si$_{1-x}$C$_{x}$ alloys show significant degradation in electron mobility in response to C incorporation. Theoretical analysis by Vaughan et al.~has demonstrated that the electron mobility in dilute Si$_{1-x}$C$_{x}$ is limited not only by C-related alloy scattering, but also by electrically active crystalline defects. Based on our electronic structure analysis here, we similarly expect a strong C-induced reduction of the electron mobility in dilute Ge$_{1-x}$C$_{x}$ alloys. The identification in our disordered supercell calculations of strong inhomogeneous energetic broadening of the CB edge Bloch character reflects the presence of a distribution of localised states and indicates a breakdown of strict \textbf{k}-selection. On this basis, we expect many scattering pathways to become available for electrons in the Ge$_{1-x}$C$_{x}$ CB, so that the electron mobility in the alloy is both strongly and intrinsically limited.

We emphasise that the conclusions drawn above are based solely on the results of our analysis of the electronic structure of idealised Ge$_{1-x}$C$_{x}$ alloys -- i.e.~purely substitutional, defect-free alloys. The presence of defects -- e.g. in the form of C interstitials due to non-substitutional incorporation -- will likely exacerbate limitations on the optical and transport properties. Overall, while dilute Ge$_{1-x}$C$_{x}$ alloys are of interest from a fundamental perspective due to their unusual electronic properties, we conclude that they are likely to be of limited value for applications in CMOS-compatible optoelectronic devices.

%%%%%%%%%%%%%%%%%%%%%
%%%% Conclusions %%%%
%%%%%%%%%%%%%%%%%%%%%

\section{Conclusions}
\label{sec:conclusions}

% Summary

In summary, we have presented a theoretical analysis of electronic structure evolution in the group-IV dilute carbide alloy Ge$_{1-x}$C$_{x}$. Our calculations were based on a computationally efficient and highly-scalable semi-empirical atomistic framework, consisting of (i) structural relaxation using a VFF potential, and (ii) electronic structure calculations using a $sp^{3}s^{\ast}$ TB Hamiltonian. Both the VFF potential and TB Hamiltonian were parametrised based on hybrid functional DFT calculations of the structural, elastic and electronic properties of the constituent diamond-structured elemental semiconductors Ge and C, as well as the IV-IV compound semiconductor zb-GeC. The validity of this framework has been established by comparison to the results of hybrid functional DFT calculations for small alloy supercells containing up to 128 atoms, where good qualitative and quantitative agreement is found for the nature and evolution of the alloy band gap. Using this framework we investigated the evolution with C composition $x$ of the electronic structure of idealised (ordered) and realistic (disordered) Ge$_{1-x}$C$_{x}$ alloys.

% What we did

Recently, it has been suggested that C incorporation in Ge drives the evolution of a direct band gap via the formation of C-related localised impurity states, with these C-related states undergoing a BAC interaction with the extended $\Gamma_{7c}$ zone centre CB edge states of the Ge host matrix semiconductor. Contrary to this suggestion, our calculations for ordered Ge$_{1-x}$C$_{x}$ alloys revealed the presence of weak C-induced mixing of Ge $\Gamma_{7c}$ character into the alloy CB edge state. As such, rather than being formed via a BAC interaction of an admixture of a C-related localised impurity state, we showed that the CB edge in ordered Ge$_{1-x}$C$_{x}$ alloy supercells is predominantly formed of an $A_{1}$-symmetric ($s$-like) linear combination of the extended L$_{6c}$ CB edge states of Ge. Consequently, we demonstrated that the lowest energy CB state in ordered Ge$_{1-x}$C$_{x}$ supercells displays minimal localisation about C lattice sites as the ultra-dilute limit is approached.

For large, disordered Ge$_{1-x}$C$_{x}$ alloy supercells we calculated the evolution of the electronic structure up to $x = 2$\%. Generally, we found that the lowest energy Ge$_{1-x}$C$_{x}$ alloy CB states acquire only a minimal admixture of Ge $\Gamma_{7c}$ (direct-gap) character, while retaining predominantly Ge L$_{6c}$ (indirect-gap) character. With increasing $x$ our calculations revealed that C-related alloy disorder leads to strong inhomogeneous energetic broadening of the Bloch character associated with the CB edge states, due to the formation of a distribution of localised states, associated with C clustering and lying energetically within the Ge band gap. The formation of a distribution of localised states within the Ge band gap indicates that dilute Ge$_{1-x}$C$_{x}$ alloys are likely to possess intrinsically poor optical and transport properties. On average, we calculated that the direct (Ge $\Gamma_{7c}$) character becomes distributed across a multiplicity of higher energy alloy CB states with increasing C composition $x$, behaviour which is markedly different to that expected based on the BAC model.

% General conclusions

In conclusion, rather than acquiring a direct band gap via a BAC interaction, our analysis demonstrates that the Ge$_{1-x}$C$_{x}$ CB edge attains minimal direct character. The C-induced band mixing identified by our calculations instead leads to the formation of a ``quasi-direct'' hybridised alloy band gap in the dilute C limit, which retains predominantly indirect (Ge L$_{6c}$) character, with a distribution of C localised states emerging within the Ge band gap in the presence of short-range alloy disorder. These general conclusions are in qualitative and quantitative agreement with those of recent analysis based on hybrid functional DFT calculations. \cite{Kirwan_SST_2019} We therefore conclude that C incorporation in Ge does \textit{not} give rise to a direct-gap semiconductor alloy, limiting the potential of this material system for applications in optoelectronic devices.

%%%%%%%%%%%%%%%%%%%%%%%%%%
%%%% Acknowledgements %%%%
%%%%%%%%%%%%%%%%%%%%%%%%%%

\section*{Acknowledgements}

This work was supported by Science Foundation Ireland (SFI; project no.~15/IA/3082), and by the National University of Ireland (NUI; via the Post-Doctoral Fellowship in the Sciences, held by C.A.B.). The authors thank Ms.~Amy C.~Kirwan and Dr.~Stefan Schulz (Tyndall National Institute, Ireland) for useful discussions, and for providing access to the results of their theoretical calculations prior to publication.

%%%%%%%%%%%%%%%%%%%%%%%%%%%%%%%
%%%% Data access statement %%%%
%%%%%%%%%%%%%%%%%%%%%%%%%%%%%%%

% \section*{Data access}

% The data associated with this work are openly available, and can be accessed via Ref.~\onlinecite{Broderick_GeC_2019_data}.

%%%%%%%%%%%%%%%%%%%%
%%%% References %%%%
%%%%%%%%%%%%%%%%%%%%

% \bibliographystyle{apsrev}                   % Using the Physical Review style for references
% \bibliography     {GeC_electronic_structure} % The bibliography is contained in the file: GeC_electronic_structure.bib

\begin{thebibliography}{84}
\expandafter\ifx\csname natexlab\endcsname\relax\def\natexlab#1{#1}\fi
\expandafter\ifx\csname bibnamefont\endcsname\relax
  \def\bibnamefont#1{#1}\fi
\expandafter\ifx\csname bibfnamefont\endcsname\relax
  \def\bibfnamefont#1{#1}\fi
\expandafter\ifx\csname citenamefont\endcsname\relax
  \def\citenamefont#1{#1}\fi
\expandafter\ifx\csname url\endcsname\relax
  \def\url#1{\texttt{#1}}\fi
\expandafter\ifx\csname urlprefix\endcsname\relax\def\urlprefix{URL }\fi
\providecommand{\bibinfo}[2]{#2}
\providecommand{\eprint}[2][]{\url{#2}}

\bibitem[{\citenamefont{Zhou et~al.}(2015)\citenamefont{Zhou, Yin, and
  Michel}}]{Zhou_LSAA_2015}
\bibinfo{author}{\bibfnamefont{Z.}~\bibnamefont{Zhou}},
  \bibinfo{author}{\bibfnamefont{B.}~\bibnamefont{Yin}}, \bibnamefont{and}
  \bibinfo{author}{\bibfnamefont{J.}~\bibnamefont{Michel}},
  \bibinfo{journal}{Light: Science \& Applications}
  \textbf{\bibinfo{volume}{4}}, \bibinfo{pages}{e358} (\bibinfo{year}{2015}).

\bibitem[{\citenamefont{Saito et~al.}(2016)\citenamefont{Saito, Al-Attili, Oda,
  and Ishikawa}}]{Saito_SST_2016}
\bibinfo{author}{\bibfnamefont{S.}~\bibnamefont{Saito}},
  \bibinfo{author}{\bibfnamefont{A.~Z.} \bibnamefont{Al-Attili}},
  \bibinfo{author}{\bibfnamefont{K.}~\bibnamefont{Oda}}, \bibnamefont{and}
  \bibinfo{author}{\bibfnamefont{Y.}~\bibnamefont{Ishikawa}},
  \bibinfo{journal}{Semicond.~Sci.~Technol.} \textbf{\bibinfo{volume}{31}},
  \bibinfo{pages}{043002} (\bibinfo{year}{2016}).

\bibitem[{\citenamefont{Thomson et~al.}(2016)\citenamefont{Thomson, Zilkie,
  Bowers, Komljenovic, Reed, Vivien, Marris-Morini, Cassan, Virot,
  F{\'{e}}d{\'{e}}li et~al.}}]{Thomson_JO_2016}
\bibinfo{author}{\bibfnamefont{D.}~\bibnamefont{Thomson}},
  \bibinfo{author}{\bibfnamefont{A.}~\bibnamefont{Zilkie}},
  \bibinfo{author}{\bibfnamefont{J.~E.} \bibnamefont{Bowers}},
  \bibinfo{author}{\bibfnamefont{T.}~\bibnamefont{Komljenovic}},
  \bibinfo{author}{\bibfnamefont{G.~T.} \bibnamefont{Reed}},
  \bibinfo{author}{\bibfnamefont{L.}~\bibnamefont{Vivien}},
  \bibinfo{author}{\bibfnamefont{D.}~\bibnamefont{Marris-Morini}},
  \bibinfo{author}{\bibfnamefont{E.}~\bibnamefont{Cassan}},
  \bibinfo{author}{\bibfnamefont{L.}~\bibnamefont{Virot}},
  \bibinfo{author}{\bibfnamefont{J.-M.} \bibnamefont{F{\'{e}}d{\'{e}}li}},
  \bibnamefont{et~al.}, \bibinfo{journal}{J.~Opt.}
  \textbf{\bibinfo{volume}{18}}, \bibinfo{pages}{073003}
  (\bibinfo{year}{2016}).

\bibitem[{\citenamefont{Yamaguchi et~al.}(2005)\citenamefont{Yamaguchi,
  Takamoto, Araki, and Ekins-Daukes}}]{Yamaguchi_SE_2005}
\bibinfo{author}{\bibfnamefont{M.}~\bibnamefont{Yamaguchi}},
  \bibinfo{author}{\bibfnamefont{T.}~\bibnamefont{Takamoto}},
  \bibinfo{author}{\bibfnamefont{K.}~\bibnamefont{Araki}}, \bibnamefont{and}
  \bibinfo{author}{\bibfnamefont{N.~J.} \bibnamefont{Ekins-Daukes}},
  \bibinfo{journal}{Solar Energy} \textbf{\bibinfo{volume}{79}},
  \bibinfo{pages}{78} (\bibinfo{year}{2005}).

\bibitem[{\citenamefont{Yamaguchi et~al.}(2008)\citenamefont{Yamaguchi,
  Nishimura, Sasaki, Suzuki, Arafune, Kojima, Ohsita, Okada, Yamamoto, Takamoto
  et~al.}}]{Yamaguchi_SE_2008}
\bibinfo{author}{\bibfnamefont{M.}~\bibnamefont{Yamaguchi}},
  \bibinfo{author}{\bibfnamefont{K.-I.} \bibnamefont{Nishimura}},
  \bibinfo{author}{\bibfnamefont{T.}~\bibnamefont{Sasaki}},
  \bibinfo{author}{\bibfnamefont{H.}~\bibnamefont{Suzuki}},
  \bibinfo{author}{\bibfnamefont{K.}~\bibnamefont{Arafune}},
  \bibinfo{author}{\bibfnamefont{N.}~\bibnamefont{Kojima}},
  \bibinfo{author}{\bibfnamefont{Y.}~\bibnamefont{Ohsita}},
  \bibinfo{author}{\bibfnamefont{Y.}~\bibnamefont{Okada}},
  \bibinfo{author}{\bibfnamefont{A.}~\bibnamefont{Yamamoto}},
  \bibinfo{author}{\bibfnamefont{T.}~\bibnamefont{Takamoto}},
  \bibnamefont{et~al.}, \bibinfo{journal}{Solar Energy}
  \textbf{\bibinfo{volume}{82}}, \bibinfo{pages}{173} (\bibinfo{year}{2008}).

\bibitem[{\citenamefont{Roucka et~al.}(2016)\citenamefont{Roucka, Clark,
  Wilson, Thomas, F{\"{u}}hrer, Ekins-Daukes, Johnson, Hoffman, and
  Begarney}}]{Roucka_IEEEJPV_2016}
\bibinfo{author}{\bibfnamefont{R.}~\bibnamefont{Roucka}},
  \bibinfo{author}{\bibfnamefont{A.}~\bibnamefont{Clark}},
  \bibinfo{author}{\bibfnamefont{T.}~\bibnamefont{Wilson}},
  \bibinfo{author}{\bibfnamefont{T.}~\bibnamefont{Thomas}},
  \bibinfo{author}{\bibfnamefont{M.}~\bibnamefont{F{\"{u}}hrer}},
  \bibinfo{author}{\bibfnamefont{N.~J.} \bibnamefont{Ekins-Daukes}},
  \bibinfo{author}{\bibfnamefont{A.}~\bibnamefont{Johnson}},
  \bibinfo{author}{\bibfnamefont{R.}~\bibnamefont{Hoffman}}, \bibnamefont{and}
  \bibinfo{author}{\bibfnamefont{D.}~\bibnamefont{Begarney}},
  \bibinfo{journal}{IEEE J.~Photovolt.} \textbf{\bibinfo{volume}{6}},
  \bibinfo{pages}{1025} (\bibinfo{year}{2016}).

\bibitem[{\citenamefont{Geiger et~al.}(2015)\citenamefont{Geiger, Zabel, and
  Sigg}}]{Geiger_FM_2015}
\bibinfo{author}{\bibfnamefont{R.}~\bibnamefont{Geiger}},
  \bibinfo{author}{\bibfnamefont{T.}~\bibnamefont{Zabel}}, \bibnamefont{and}
  \bibinfo{author}{\bibfnamefont{H.}~\bibnamefont{Sigg}},
  \bibinfo{journal}{Front.~Mater.} \textbf{\bibinfo{volume}{2}},
  \bibinfo{pages}{52} (\bibinfo{year}{2015}).

\bibitem[{\citenamefont{Reboud et~al.}(2017)\citenamefont{Reboud, Gassenq,
  Hartmann, Widiez, Virot, Aubin, Guilloy, Tardif, F{\'{e}}d{\'{e}}li, Pauc
  et~al.}}]{Reboud_PCGC_2017}
\bibinfo{author}{\bibfnamefont{V.}~\bibnamefont{Reboud}},
  \bibinfo{author}{\bibfnamefont{A.}~\bibnamefont{Gassenq}},
  \bibinfo{author}{\bibfnamefont{J.~M.} \bibnamefont{Hartmann}},
  \bibinfo{author}{\bibfnamefont{J.}~\bibnamefont{Widiez}},
  \bibinfo{author}{\bibfnamefont{L.}~\bibnamefont{Virot}},
  \bibinfo{author}{\bibfnamefont{J.}~\bibnamefont{Aubin}},
  \bibinfo{author}{\bibfnamefont{K.}~\bibnamefont{Guilloy}},
  \bibinfo{author}{\bibfnamefont{S.}~\bibnamefont{Tardif}},
  \bibinfo{author}{\bibfnamefont{J.~M.} \bibnamefont{F{\'{e}}d{\'{e}}li}},
  \bibinfo{author}{\bibfnamefont{N.}~\bibnamefont{Pauc}}, \bibnamefont{et~al.},
  \bibinfo{journal}{Prog.~Cryst.~Growth \& Charact.}
  \textbf{\bibinfo{volume}{63}}, \bibinfo{pages}{1} (\bibinfo{year}{2017}).

\bibitem[{\citenamefont{Zhang et~al.}(2009)\citenamefont{Zhang, Crespi, and
  Zhang}}]{Zhang_PRL_2009}
\bibinfo{author}{\bibfnamefont{F.}~\bibnamefont{Zhang}},
  \bibinfo{author}{\bibfnamefont{V.~H.} \bibnamefont{Crespi}},
  \bibnamefont{and} \bibinfo{author}{\bibfnamefont{P.}~\bibnamefont{Zhang}},
  \bibinfo{journal}{Phys.~Rev.~Lett.} \textbf{\bibinfo{volume}{102}},
  \bibinfo{pages}{156401} (\bibinfo{year}{2009}).

\bibitem[{\citenamefont{Kurdia et~al.}(2010)\citenamefont{Kurdia, Bertin,
  Martincic, de~Kersauson, Fishman, Sauvage, Bosseboeuf, and
  Boucauda}}]{Kurdia_APL_2010}
\bibinfo{author}{\bibfnamefont{M.~E.} \bibnamefont{Kurdia}},
  \bibinfo{author}{\bibfnamefont{H.}~\bibnamefont{Bertin}},
  \bibinfo{author}{\bibfnamefont{E.}~\bibnamefont{Martincic}},
  \bibinfo{author}{\bibfnamefont{M.}~\bibnamefont{de~Kersauson}},
  \bibinfo{author}{\bibfnamefont{G.}~\bibnamefont{Fishman}},
  \bibinfo{author}{\bibfnamefont{S.}~\bibnamefont{Sauvage}},
  \bibinfo{author}{\bibfnamefont{A.}~\bibnamefont{Bosseboeuf}},
  \bibnamefont{and} \bibinfo{author}{\bibfnamefont{P.}~\bibnamefont{Boucauda}},
  \bibinfo{journal}{Appl.~Phys.~Lett.} \textbf{\bibinfo{volume}{96}},
  \bibinfo{pages}{041909} (\bibinfo{year}{2010}).

\bibitem[{\citenamefont{S\'{a}nchez-P\'{e}rez
  et~al.}(2011)\citenamefont{S\'{a}nchez-P\'{e}rez, Boztug, Chen, Sudradjat,
  Paskiewicz, Jacobson, Lagally, and Paiella}}]{Sanchez-Perez_PNAS_2011}
\bibinfo{author}{\bibfnamefont{J.~R.} \bibnamefont{S\'{a}nchez-P\'{e}rez}},
  \bibinfo{author}{\bibfnamefont{C.}~\bibnamefont{Boztug}},
  \bibinfo{author}{\bibfnamefont{F.}~\bibnamefont{Chen}},
  \bibinfo{author}{\bibfnamefont{F.~F.} \bibnamefont{Sudradjat}},
  \bibinfo{author}{\bibfnamefont{D.~M.} \bibnamefont{Paskiewicz}},
  \bibinfo{author}{\bibfnamefont{R.~B.} \bibnamefont{Jacobson}},
  \bibinfo{author}{\bibfnamefont{M.~G.} \bibnamefont{Lagally}},
  \bibnamefont{and} \bibinfo{author}{\bibfnamefont{R.}~\bibnamefont{Paiella}},
  \bibinfo{journal}{Proc.~Natl.~Acad.~Sci.~U.S.A.}
  \textbf{\bibinfo{volume}{108}}, \bibinfo{pages}{18893}
  (\bibinfo{year}{2011}).

\bibitem[{\citenamefont{S\"{u}ess et~al.}(2013)\citenamefont{S\"{u}ess, Geiger,
  Minamisawa, Schiefler, Frigerio, Chrastina, Isella, Spolenak, Faist, and
  Sigg}}]{Suess_NP_2013}
\bibinfo{author}{\bibfnamefont{M.~J.} \bibnamefont{S\"{u}ess}},
  \bibinfo{author}{\bibfnamefont{R.}~\bibnamefont{Geiger}},
  \bibinfo{author}{\bibfnamefont{R.~A.} \bibnamefont{Minamisawa}},
  \bibinfo{author}{\bibfnamefont{G.}~\bibnamefont{Schiefler}},
  \bibinfo{author}{\bibfnamefont{J.}~\bibnamefont{Frigerio}},
  \bibinfo{author}{\bibfnamefont{D.}~\bibnamefont{Chrastina}},
  \bibinfo{author}{\bibfnamefont{G.}~\bibnamefont{Isella}},
  \bibinfo{author}{\bibfnamefont{R.}~\bibnamefont{Spolenak}},
  \bibinfo{author}{\bibfnamefont{J.}~\bibnamefont{Faist}}, \bibnamefont{and}
  \bibinfo{author}{\bibfnamefont{H.}~\bibnamefont{Sigg}},
  \bibinfo{journal}{Nature Photonics} \textbf{\bibinfo{volume}{7}},
  \bibinfo{pages}{466} (\bibinfo{year}{2013}).

\bibitem[{\citenamefont{Kouvetakis et~al.}(2006)\citenamefont{Kouvetakis,
  Menendez, and Chizmeshya}}]{Kouvetakis_ARMR_2006}
\bibinfo{author}{\bibfnamefont{J.}~\bibnamefont{Kouvetakis}},
  \bibinfo{author}{\bibfnamefont{J.}~\bibnamefont{Menendez}}, \bibnamefont{and}
  \bibinfo{author}{\bibfnamefont{A.~V.~G.} \bibnamefont{Chizmeshya}},
  \bibinfo{journal}{Annu.~Rev.~Mater.~Res.} \textbf{\bibinfo{volume}{36}},
  \bibinfo{pages}{497} (\bibinfo{year}{2006}).

\bibitem[{\citenamefont{Moontragoon et~al.}(2007)\citenamefont{Moontragoon,
  Ikoni\'{c}, and Harrison}}]{Moontragoon_SST_2007}
\bibinfo{author}{\bibfnamefont{P.}~\bibnamefont{Moontragoon}},
  \bibinfo{author}{\bibfnamefont{Z.}~\bibnamefont{Ikoni\'{c}}},
  \bibnamefont{and} \bibinfo{author}{\bibfnamefont{P.}~\bibnamefont{Harrison}},
  \bibinfo{journal}{Semicond.~Sci.~Technol.} \textbf{\bibinfo{volume}{22}},
  \bibinfo{pages}{742} (\bibinfo{year}{2007}).

\bibitem[{\citenamefont{Soref}(2014)}]{Soref_PTRSA_2014}
\bibinfo{author}{\bibfnamefont{R.}~\bibnamefont{Soref}},
  \bibinfo{journal}{Phil.~Trans.~R.~Soc.~A} \textbf{\bibinfo{volume}{372}},
  \bibinfo{pages}{0113} (\bibinfo{year}{2014}).

\bibitem[{\citenamefont{Zaima et~al.}(2015)\citenamefont{Zaima, Nakatsuka,
  Taoka, Kurosawa, Takeuchi, and Sakashita}}]{Zaima_STAM_2015}
\bibinfo{author}{\bibfnamefont{S.}~\bibnamefont{Zaima}},
  \bibinfo{author}{\bibfnamefont{O.}~\bibnamefont{Nakatsuka}},
  \bibinfo{author}{\bibfnamefont{N.}~\bibnamefont{Taoka}},
  \bibinfo{author}{\bibfnamefont{M.}~\bibnamefont{Kurosawa}},
  \bibinfo{author}{\bibfnamefont{W.}~\bibnamefont{Takeuchi}}, \bibnamefont{and}
  \bibinfo{author}{\bibfnamefont{M.}~\bibnamefont{Sakashita}},
  \bibinfo{journal}{Sci.~Technol.~Adv.~Mater.} \textbf{\bibinfo{volume}{16}},
  \bibinfo{pages}{043502} (\bibinfo{year}{2015}).

\bibitem[{\citenamefont{Wirths et~al.}(2015)\citenamefont{Wirths, Geiger,
  von~den Driesch, Mussler, Stoica, Mantl, Ikonic, Luysberg, Chiussi, Hartmann
  et~al.}}]{Wirths_NP_2015}
\bibinfo{author}{\bibfnamefont{S.}~\bibnamefont{Wirths}},
  \bibinfo{author}{\bibfnamefont{R.}~\bibnamefont{Geiger}},
  \bibinfo{author}{\bibfnamefont{N.}~\bibnamefont{von~den Driesch}},
  \bibinfo{author}{\bibfnamefont{G.}~\bibnamefont{Mussler}},
  \bibinfo{author}{\bibfnamefont{T.}~\bibnamefont{Stoica}},
  \bibinfo{author}{\bibfnamefont{S.}~\bibnamefont{Mantl}},
  \bibinfo{author}{\bibfnamefont{Z.}~\bibnamefont{Ikonic}},
  \bibinfo{author}{\bibfnamefont{M.}~\bibnamefont{Luysberg}},
  \bibinfo{author}{\bibfnamefont{S.}~\bibnamefont{Chiussi}},
  \bibinfo{author}{\bibfnamefont{J.~M.} \bibnamefont{Hartmann}},
  \bibnamefont{et~al.}, \bibinfo{journal}{Nat.~Photonics}
  \textbf{\bibinfo{volume}{9}}, \bibinfo{pages}{88} (\bibinfo{year}{2015}).

\bibitem[{\citenamefont{{J.~Margetis, Y.~Zhou, W.~Dou, P.~C.~Grant,
  B.~Alharthi, et al.}}(2018)}]{Margetis_APL_2018}
\bibinfo{author}{\bibnamefont{{J.~Margetis, Y.~Zhou, W.~Dou, P.~C.~Grant,
  B.~Alharthi, et al.}}}, \bibinfo{journal}{Appl.~Phys.~Lett.}
  \textbf{\bibinfo{volume}{113}}, \bibinfo{pages}{221104}
  (\bibinfo{year}{2018}).

\bibitem[{\citenamefont{Huang et~al.}(2014)\citenamefont{Huang, Cheng, Xue, and
  Li}}]{Huang_PB_2014}
\bibinfo{author}{\bibfnamefont{W.}~\bibnamefont{Huang}},
  \bibinfo{author}{\bibfnamefont{B.}~\bibnamefont{Cheng}},
  \bibinfo{author}{\bibfnamefont{C.}~\bibnamefont{Xue}}, \bibnamefont{and}
  \bibinfo{author}{\bibfnamefont{C.}~\bibnamefont{Li}},
  \bibinfo{journal}{Physica B} \textbf{\bibinfo{volume}{443}},
  \bibinfo{pages}{43} (\bibinfo{year}{2014}).

\bibitem[{\citenamefont{Huang et~al.}(2017)\citenamefont{Huang, Cheng, Xue, and
  Yang}}]{Huang_JAC_2017}
\bibinfo{author}{\bibfnamefont{W.}~\bibnamefont{Huang}},
  \bibinfo{author}{\bibfnamefont{B.}~\bibnamefont{Cheng}},
  \bibinfo{author}{\bibfnamefont{C.}~\bibnamefont{Xue}}, \bibnamefont{and}
  \bibinfo{author}{\bibfnamefont{H.}~\bibnamefont{Yang}},
  \bibinfo{journal}{J.~Alloy Compd.} \textbf{\bibinfo{volume}{701}},
  \bibinfo{pages}{816} (\bibinfo{year}{2017}).

\bibitem[{\citenamefont{Alahmad et~al.}(2018)\citenamefont{Alahmad, Mosleh,
  Alher, Fahimeh, Banihashemian, Ghetmiri, Al-Kabi, Du, Li, Yu
  et~al.}}]{Alahmad_JEM_2018}
\bibinfo{author}{\bibfnamefont{H.}~\bibnamefont{Alahmad}},
  \bibinfo{author}{\bibfnamefont{A.}~\bibnamefont{Mosleh}},
  \bibinfo{author}{\bibfnamefont{M.}~\bibnamefont{Alher}},
  \bibinfo{author}{\bibfnamefont{S.}~\bibnamefont{Fahimeh}},
  \bibinfo{author}{\bibnamefont{Banihashemian}},
  \bibinfo{author}{\bibfnamefont{S.~A.} \bibnamefont{Ghetmiri}},
  \bibinfo{author}{\bibfnamefont{S.}~\bibnamefont{Al-Kabi}},
  \bibinfo{author}{\bibfnamefont{W.}~\bibnamefont{Du}},
  \bibinfo{author}{\bibfnamefont{B.}~\bibnamefont{Li}},
  \bibinfo{author}{\bibfnamefont{S.-Q.} \bibnamefont{Yu}},
  \bibnamefont{et~al.}, \bibinfo{journal}{J.~Electron.~Mater.}
  \textbf{\bibinfo{volume}{47}}, \bibinfo{pages}{3733} (\bibinfo{year}{2018}).

\bibitem[{\citenamefont{Liu et~al.}(2019)\citenamefont{Liu, Zheng, Zhou, Liu,
  Zuo, Xueab, and Cheng}}]{Liu_JAC_2019}
\bibinfo{author}{\bibfnamefont{X.}~\bibnamefont{Liu}},
  \bibinfo{author}{\bibfnamefont{J.}~\bibnamefont{Zheng}},
  \bibinfo{author}{\bibfnamefont{L.}~\bibnamefont{Zhou}},
  \bibinfo{author}{\bibfnamefont{Z.}~\bibnamefont{Liu}},
  \bibinfo{author}{\bibfnamefont{Y.}~\bibnamefont{Zuo}},
  \bibinfo{author}{\bibfnamefont{C.}~\bibnamefont{Xueab}}, \bibnamefont{and}
  \bibinfo{author}{\bibfnamefont{B.}~\bibnamefont{Cheng}},
  \bibinfo{journal}{J.~Alloy Compd.} \textbf{\bibinfo{volume}{785}},
  \bibinfo{pages}{228} (\bibinfo{year}{2019}).

\bibitem[{\citenamefont{Broderick
  et~al.}(2019{\natexlab{a}})\citenamefont{Broderick, O'Halloran, and
  O'Reilly}}]{Broderick_GePb_2019}
\bibinfo{author}{\bibfnamefont{C.~A.} \bibnamefont{Broderick}},
  \bibinfo{author}{\bibfnamefont{E.~J.} \bibnamefont{O'Halloran}},
  \bibnamefont{and} \bibinfo{author}{\bibfnamefont{E.~P.}
  \bibnamefont{O'Reilly}}, \bibinfo{journal}{{``\textit{First principles
  analysis of electronic structure evolution and the indirect- to direct-gap
  transition in Ge$_{1-x}$Pb$_{x}$ group-IV alloys}''},
  \texttt{arXiv:1911.05679}}  (\bibinfo{year}{2019}{\natexlab{a}}).

\bibitem[{\citenamefont{Stephenson
  et~al.}(2016{\natexlab{a}})\citenamefont{Stephenson, O'Brien, Penninger,
  Schneider, Gillett-Kunnath, Zajicek, Yu, Kudraweic, Stillwell, and
  Wistey}}]{Stephenson_JAP_2016}
\bibinfo{author}{\bibfnamefont{C.~A.} \bibnamefont{Stephenson}},
  \bibinfo{author}{\bibfnamefont{W.~A.} \bibnamefont{O'Brien}},
  \bibinfo{author}{\bibfnamefont{M.~W.} \bibnamefont{Penninger}},
  \bibinfo{author}{\bibfnamefont{W.~F.} \bibnamefont{Schneider}},
  \bibinfo{author}{\bibfnamefont{M.}~\bibnamefont{Gillett-Kunnath}},
  \bibinfo{author}{\bibfnamefont{J.}~\bibnamefont{Zajicek}},
  \bibinfo{author}{\bibfnamefont{K.~M.} \bibnamefont{Yu}},
  \bibinfo{author}{\bibfnamefont{R.}~\bibnamefont{Kudraweic}},
  \bibinfo{author}{\bibfnamefont{R.~A.} \bibnamefont{Stillwell}},
  \bibnamefont{and} \bibinfo{author}{\bibfnamefont{M.~A.}
  \bibnamefont{Wistey}}, \bibinfo{journal}{J.~Appl.~Phys.}
  \textbf{\bibinfo{volume}{120}}, \bibinfo{pages}{053102}
  (\bibinfo{year}{2016}{\natexlab{a}}).

\bibitem[{\citenamefont{Stephenson
  et~al.}(2016{\natexlab{b}})\citenamefont{Stephenson, O'Brien, Qi, Penninger,
  Schneider, and Wistey}}]{Stephenson_JEM_2016}
\bibinfo{author}{\bibfnamefont{C.~A.} \bibnamefont{Stephenson}},
  \bibinfo{author}{\bibfnamefont{W.~A.} \bibnamefont{O'Brien}},
  \bibinfo{author}{\bibfnamefont{M.}~\bibnamefont{Qi}},
  \bibinfo{author}{\bibfnamefont{M.~W.} \bibnamefont{Penninger}},
  \bibinfo{author}{\bibfnamefont{W.~F.} \bibnamefont{Schneider}},
  \bibnamefont{and} \bibinfo{author}{\bibfnamefont{M.~A.}
  \bibnamefont{Wistey}}, \bibinfo{journal}{J.~Electron.~Mater.}
  \textbf{\bibinfo{volume}{45}}, \bibinfo{pages}{2121}
  (\bibinfo{year}{2016}{\natexlab{b}}).

\bibitem[{\citenamefont{Broderick et~al.}(2018)\citenamefont{Broderick, Dunne,
  Tanner, Kirwan, O'Halloran, Schulz, and O'Reilly}}]{Broderick_IEEENano_2018}
\bibinfo{author}{\bibfnamefont{C.~A.} \bibnamefont{Broderick}},
  \bibinfo{author}{\bibfnamefont{M.~D.} \bibnamefont{Dunne}},
  \bibinfo{author}{\bibfnamefont{D.~S.~P.} \bibnamefont{Tanner}},
  \bibinfo{author}{\bibfnamefont{A.~C.} \bibnamefont{Kirwan}},
  \bibinfo{author}{\bibfnamefont{E.~J.} \bibnamefont{O'Halloran}},
  \bibinfo{author}{\bibfnamefont{S.}~\bibnamefont{Schulz}}, \bibnamefont{and}
  \bibinfo{author}{\bibfnamefont{E.~P.} \bibnamefont{O'Reilly}},
  \bibinfo{journal}{Proc.~18$^{\textrm{th}}$ IEEE International Conference on
  Nanotechnology}  (\bibinfo{year}{2018}).

\bibitem[{\citenamefont{Kirwan et~al.}(2019)\citenamefont{Kirwan, Schulz, and
  O'Reilly}}]{Kirwan_SST_2019}
\bibinfo{author}{\bibfnamefont{A.~C.} \bibnamefont{Kirwan}},
  \bibinfo{author}{\bibfnamefont{S.}~\bibnamefont{Schulz}}, \bibnamefont{and}
  \bibinfo{author}{\bibfnamefont{E.~P.} \bibnamefont{O'Reilly}},
  \bibinfo{journal}{Semicond.~Sci.~Technol.} \textbf{\bibinfo{volume}{34}},
  \bibinfo{pages}{075007} (\bibinfo{year}{2019}).

\bibitem[{\citenamefont{Osten et~al.}(1995)\citenamefont{Osten, Dietrich,
  R\"{u}cker, and Methfessel}}]{Osten_JCG_1995}
\bibinfo{author}{\bibfnamefont{H.~J.} \bibnamefont{Osten}},
  \bibinfo{author}{\bibfnamefont{B.}~\bibnamefont{Dietrich}},
  \bibinfo{author}{\bibfnamefont{H.}~\bibnamefont{R\"{u}cker}},
  \bibnamefont{and}
  \bibinfo{author}{\bibfnamefont{M.}~\bibnamefont{Methfessel}},
  \bibinfo{journal}{J.~Cryst.~Growth} \textbf{\bibinfo{volume}{150}},
  \bibinfo{pages}{931} (\bibinfo{year}{1995}).

\bibitem[{\citenamefont{Jain et~al.}(1995)\citenamefont{Jain, Osten, Dietrich,
  and R\"{u}cker}}]{Jain_SST_1995}
\bibinfo{author}{\bibfnamefont{S.~C.} \bibnamefont{Jain}},
  \bibinfo{author}{\bibfnamefont{H.~J.} \bibnamefont{Osten}},
  \bibinfo{author}{\bibfnamefont{B.}~\bibnamefont{Dietrich}}, \bibnamefont{and}
  \bibinfo{author}{\bibfnamefont{H.}~\bibnamefont{R\"{u}cker}},
  \bibinfo{journal}{Semicond.~Sci.~Technol.} \textbf{\bibinfo{volume}{10}},
  \bibinfo{pages}{1289} (\bibinfo{year}{1995}).

\bibitem[{\citenamefont{Finkman et~al.}(2001)\citenamefont{Finkman, Meyer, and
  Mamor}}]{Finkman_JAP_2001}
\bibinfo{author}{\bibfnamefont{E.}~\bibnamefont{Finkman}},
  \bibinfo{author}{\bibfnamefont{F.}~\bibnamefont{Meyer}}, \bibnamefont{and}
  \bibinfo{author}{\bibfnamefont{M.}~\bibnamefont{Mamor}},
  \bibinfo{journal}{J.~Appl.~Phys.} \textbf{\bibinfo{volume}{89}},
  \bibinfo{pages}{2580} (\bibinfo{year}{2001}).

\bibitem[{\citenamefont{R\"{u}cker et~al.}(1996)\citenamefont{R\"{u}cker,
  Methfessel, Dietrich, Pressel, and Osten}}]{Rucker_PRB_1996}
\bibinfo{author}{\bibfnamefont{H.}~\bibnamefont{R\"{u}cker}},
  \bibinfo{author}{\bibfnamefont{M.}~\bibnamefont{Methfessel}},
  \bibinfo{author}{\bibfnamefont{B.}~\bibnamefont{Dietrich}},
  \bibinfo{author}{\bibfnamefont{K.}~\bibnamefont{Pressel}}, \bibnamefont{and}
  \bibinfo{author}{\bibfnamefont{H.~J.} \bibnamefont{Osten}},
  \bibinfo{journal}{Phys.~Rev.~B} \textbf{\bibinfo{volume}{53}},
  \bibinfo{pages}{1302} (\bibinfo{year}{1996}).

\bibitem[{\citenamefont{Vaughan et~al.}(2012)\citenamefont{Vaughan,
  Murphy-Armando, and Fahy}}]{Vaughan_PRB_2012}
\bibinfo{author}{\bibfnamefont{M.~P.} \bibnamefont{Vaughan}},
  \bibinfo{author}{\bibfnamefont{F.}~\bibnamefont{Murphy-Armando}},
  \bibnamefont{and} \bibinfo{author}{\bibfnamefont{S.}~\bibnamefont{Fahy}},
  \bibinfo{journal}{Phys.~Rev.~B} \textbf{\bibinfo{volume}{85}},
  \bibinfo{pages}{165209} (\bibinfo{year}{2012}).

\bibitem[{\citenamefont{Okinaka et~al.}(2003)\citenamefont{Okinaka, Miyatake,
  Ohta, and Nunoshita}}]{Okinaka_JCG_2003}
\bibinfo{author}{\bibfnamefont{M.}~\bibnamefont{Okinaka}},
  \bibinfo{author}{\bibfnamefont{K.}~\bibnamefont{Miyatake}},
  \bibinfo{author}{\bibfnamefont{J.}~\bibnamefont{Ohta}}, \bibnamefont{and}
  \bibinfo{author}{\bibfnamefont{M.}~\bibnamefont{Nunoshita}},
  \bibinfo{journal}{J.~Cryst.~Growth} \textbf{\bibinfo{volume}{255}},
  \bibinfo{pages}{273} (\bibinfo{year}{2003}).

\bibitem[{\citenamefont{Roe et~al.}(1999)\citenamefont{Roe, Dashiell, Kolodzey,
  Bouchaud, and Lourtioz}}]{Roe_JVSTB_1999}
\bibinfo{author}{\bibfnamefont{K.~J.} \bibnamefont{Roe}},
  \bibinfo{author}{\bibfnamefont{M.~W.} \bibnamefont{Dashiell}},
  \bibinfo{author}{\bibfnamefont{J.}~\bibnamefont{Kolodzey}},
  \bibinfo{author}{\bibfnamefont{P.}~\bibnamefont{Bouchaud}}, \bibnamefont{and}
  \bibinfo{author}{\bibfnamefont{J.-M.} \bibnamefont{Lourtioz}},
  \bibinfo{journal}{J.~Vac.~Sci.~Technol.~B} \textbf{\bibinfo{volume}{17}},
  \bibinfo{pages}{1302} (\bibinfo{year}{1999}).

\bibitem[{\citenamefont{Kolodzey et~al.}(1995)\citenamefont{Kolodzey, Berger,
  Orner, Hits, Chen, Khan, Shao, Waite, Shah, Swann
  et~al.}}]{Kolodzey_JCG_1995}
\bibinfo{author}{\bibfnamefont{J.}~\bibnamefont{Kolodzey}},
  \bibinfo{author}{\bibfnamefont{P.~R.} \bibnamefont{Berger}},
  \bibinfo{author}{\bibfnamefont{B.~A.} \bibnamefont{Orner}},
  \bibinfo{author}{\bibfnamefont{D.}~\bibnamefont{Hits}},
  \bibinfo{author}{\bibfnamefont{F.}~\bibnamefont{Chen}},
  \bibinfo{author}{\bibfnamefont{A.}~\bibnamefont{Khan}},
  \bibinfo{author}{\bibfnamefont{X.}~\bibnamefont{Shao}},
  \bibinfo{author}{\bibfnamefont{M.~M.} \bibnamefont{Waite}},
  \bibinfo{author}{\bibfnamefont{S.~I.} \bibnamefont{Shah}},
  \bibinfo{author}{\bibfnamefont{C.~P.} \bibnamefont{Swann}},
  \bibnamefont{et~al.}, \bibinfo{journal}{J.~Cryst.~Growth}
  \textbf{\bibinfo{volume}{157}}, \bibinfo{pages}{386} (\bibinfo{year}{1995}).

\bibitem[{\citenamefont{Gall et~al.}(2000)\citenamefont{Gall, D'Arcy-Gall, and
  Greene}}]{Gall_PRB_2000}
\bibinfo{author}{\bibfnamefont{D.}~\bibnamefont{Gall}},
  \bibinfo{author}{\bibfnamefont{J.}~\bibnamefont{D'Arcy-Gall}},
  \bibnamefont{and} \bibinfo{author}{\bibfnamefont{J.~E.}
  \bibnamefont{Greene}}, \bibinfo{journal}{Phys.~Rev.~B}
  \textbf{\bibinfo{volume}{62}}, \bibinfo{pages}{R7723(R)}
  (\bibinfo{year}{2000}).

\bibitem[{\citenamefont{Park et~al.}(2002)\citenamefont{Park, D'Arcy-Gall,
  Gall, Kim, Desjardins, and Greene}}]{Park_JAP_2002}
\bibinfo{author}{\bibfnamefont{S.~Y.} \bibnamefont{Park}},
  \bibinfo{author}{\bibfnamefont{J.}~\bibnamefont{D'Arcy-Gall}},
  \bibinfo{author}{\bibfnamefont{D.}~\bibnamefont{Gall}},
  \bibinfo{author}{\bibfnamefont{Y.-W.} \bibnamefont{Kim}},
  \bibinfo{author}{\bibfnamefont{P.}~\bibnamefont{Desjardins}},
  \bibnamefont{and} \bibinfo{author}{\bibfnamefont{J.~E.}
  \bibnamefont{Greene}}, \bibinfo{journal}{J.~Appl.~Phys.}
  \textbf{\bibinfo{volume}{91}}, \bibinfo{pages}{3644} (\bibinfo{year}{2002}).

\bibitem[{\citenamefont{Stephenson
  et~al.}(2016{\natexlab{c}})\citenamefont{Stephenson, Gillett-Kunnath,
  O'Brien, Kudraweic, and Wistey}}]{Stephenson_C_2016}
\bibinfo{author}{\bibfnamefont{C.~A.} \bibnamefont{Stephenson}},
  \bibinfo{author}{\bibfnamefont{M.}~\bibnamefont{Gillett-Kunnath}},
  \bibinfo{author}{\bibfnamefont{W.~A.} \bibnamefont{O'Brien}},
  \bibinfo{author}{\bibfnamefont{R.}~\bibnamefont{Kudraweic}},
  \bibnamefont{and} \bibinfo{author}{\bibfnamefont{M.~A.}
  \bibnamefont{Wistey}}, \bibinfo{journal}{Crystals}
  \textbf{\bibinfo{volume}{6}}, \bibinfo{pages}{159}
  (\bibinfo{year}{2016}{\natexlab{c}}).

\bibitem[{\citenamefont{Shan et~al.}(1999)\citenamefont{Shan, Walukiewicz, III,
  Haller, Geisz, Friedman, Olson, and Kurtz}}]{Shan_PRL_1999}
\bibinfo{author}{\bibfnamefont{W.}~\bibnamefont{Shan}},
  \bibinfo{author}{\bibfnamefont{W.}~\bibnamefont{Walukiewicz}},
  \bibinfo{author}{\bibfnamefont{J.~W.~A.} \bibnamefont{III}},
  \bibinfo{author}{\bibfnamefont{E.~E.} \bibnamefont{Haller}},
  \bibinfo{author}{\bibfnamefont{J.~F.} \bibnamefont{Geisz}},
  \bibinfo{author}{\bibfnamefont{D.~J.} \bibnamefont{Friedman}},
  \bibinfo{author}{\bibfnamefont{J.~M.} \bibnamefont{Olson}}, \bibnamefont{and}
  \bibinfo{author}{\bibfnamefont{S.~R.} \bibnamefont{Kurtz}},
  \bibinfo{journal}{Phys.~Rev.~Lett.} \textbf{\bibinfo{volume}{82}},
  \bibinfo{pages}{1221} (\bibinfo{year}{1999}).

\bibitem[{\citenamefont{Wu et~al.}(2002)\citenamefont{Wu, Shan, and
  Walukiewicz}}]{Wu_SST_2002}
\bibinfo{author}{\bibfnamefont{J.}~\bibnamefont{Wu}},
  \bibinfo{author}{\bibfnamefont{W.}~\bibnamefont{Shan}}, \bibnamefont{and}
  \bibinfo{author}{\bibfnamefont{W.}~\bibnamefont{Walukiewicz}},
  \bibinfo{journal}{Semicond.~Sci.~Technol.} \textbf{\bibinfo{volume}{17}},
  \bibinfo{pages}{860} (\bibinfo{year}{2002}).

\bibitem[{\citenamefont{Wu et~al.}(2003)\citenamefont{Wu, Walukiewicz, Yu, III,
  Haller, Miotkowski, Ramdas, Su, Sou, Perera et~al.}}]{Wu_PRB_2007}
\bibinfo{author}{\bibfnamefont{J.}~\bibnamefont{Wu}},
  \bibinfo{author}{\bibfnamefont{W.}~\bibnamefont{Walukiewicz}},
  \bibinfo{author}{\bibfnamefont{K.~M.} \bibnamefont{Yu}},
  \bibinfo{author}{\bibfnamefont{J.~W.~A.} \bibnamefont{III}},
  \bibinfo{author}{\bibfnamefont{E.~E.} \bibnamefont{Haller}},
  \bibinfo{author}{\bibfnamefont{I.}~\bibnamefont{Miotkowski}},
  \bibinfo{author}{\bibfnamefont{A.~K.} \bibnamefont{Ramdas}},
  \bibinfo{author}{\bibfnamefont{C.-H.} \bibnamefont{Su}},
  \bibinfo{author}{\bibfnamefont{I.~K.} \bibnamefont{Sou}},
  \bibinfo{author}{\bibfnamefont{R.~C.~C.} \bibnamefont{Perera}},
  \bibnamefont{et~al.}, \bibinfo{journal}{Phys.~Rev.~B}
  \textbf{\bibinfo{volume}{67}}, \bibinfo{pages}{035207}
  (\bibinfo{year}{2003}).

\bibitem[{\citenamefont{O'Reilly et~al.}(2009)\citenamefont{O'Reilly, Lindsay,
  Klar, Polimeni, and Capizzi}}]{Reilly_SST_2009}
\bibinfo{author}{\bibfnamefont{E.~P.} \bibnamefont{O'Reilly}},
  \bibinfo{author}{\bibfnamefont{A.}~\bibnamefont{Lindsay}},
  \bibinfo{author}{\bibfnamefont{P.~J.} \bibnamefont{Klar}},
  \bibinfo{author}{\bibfnamefont{A.}~\bibnamefont{Polimeni}}, \bibnamefont{and}
  \bibinfo{author}{\bibfnamefont{M.}~\bibnamefont{Capizzi}},
  \bibinfo{journal}{Semicond.~Sci.~Technol.} \textbf{\bibinfo{volume}{24}},
  \bibinfo{pages}{033001} (\bibinfo{year}{2009}).

\bibitem[{\citenamefont{Tanner}(2017)}]{Tanner_thesis_2017}
\bibinfo{author}{\bibfnamefont{D.~S.~P.} \bibnamefont{Tanner}}, Ph.D. thesis,
  \bibinfo{school}{University College Cork}, \bibinfo{address}{Ireland}
  (\bibinfo{year}{2017}), \urlprefix\url{http://cora.ucc.ie/handle/10468/5459}.

\bibitem[{\citenamefont{Tanner et~al.}(2019{\natexlab{a}})\citenamefont{Tanner,
  Broderick, Kirwan, Schulz, and O'Reilly}}]{Tanner_elastic_2019}
\bibinfo{author}{\bibfnamefont{D.~S.~P.} \bibnamefont{Tanner}},
  \bibinfo{author}{\bibfnamefont{C.~A.} \bibnamefont{Broderick}},
  \bibinfo{author}{\bibfnamefont{A.~C.} \bibnamefont{Kirwan}},
  \bibinfo{author}{\bibfnamefont{S.}~\bibnamefont{Schulz}}, \bibnamefont{and}
  \bibinfo{author}{\bibfnamefont{E.~P.} \bibnamefont{O'Reilly}},
  \bibinfo{journal}{{``\textit{Elastic properties of elemental, compound, and
  alloyed group-IV materials: hybrid density functional theory and valence
  force field parametrisation}''}, in preparation}
  (\bibinfo{year}{2019}{\natexlab{a}}).

\bibitem[{\citenamefont{Broderick
  et~al.}(2019{\natexlab{b}})\citenamefont{Broderick, Kirwan, Schulz, and
  O'Reilly}}]{Broderick_electronic_2019}
\bibinfo{author}{\bibfnamefont{C.~A.} \bibnamefont{Broderick}},
  \bibinfo{author}{\bibfnamefont{A.~C.} \bibnamefont{Kirwan}},
  \bibinfo{author}{\bibfnamefont{S.}~\bibnamefont{Schulz}}, \bibnamefont{and}
  \bibinfo{author}{\bibfnamefont{E.~P.} \bibnamefont{O'Reilly}},
  \bibinfo{journal}{{``\textit{Electronic properties of elemental and compound
  group-IV materials: hybrid density functional theory and tight-binding
  parametrisation}''}, in preparation}  (\bibinfo{year}{2019}{\natexlab{b}}).

\bibitem[{\citenamefont{Martin}(1970)}]{Martin_PRB_1970}
\bibinfo{author}{\bibfnamefont{R.~M.} \bibnamefont{Martin}},
  \bibinfo{journal}{Phys.~Rev.~B} \textbf{\bibinfo{volume}{1}},
  \bibinfo{pages}{4005} (\bibinfo{year}{1970}).

\bibitem[{\citenamefont{Musgrave and Pople}(1962)}]{Musgrave_PRSLA_1962}
\bibinfo{author}{\bibfnamefont{M.~J.~P.} \bibnamefont{Musgrave}}
  \bibnamefont{and} \bibinfo{author}{\bibfnamefont{J.~A.} \bibnamefont{Pople}},
  \bibinfo{journal}{Proc.~R.~Soc.~Lond.~A} \textbf{\bibinfo{volume}{268}},
  \bibinfo{pages}{474} (\bibinfo{year}{1962}).

\bibitem[{\citenamefont{Tanner et~al.}(2019{\natexlab{b}})\citenamefont{Tanner,
  Caro, Schulz, and O'Reilly}}]{Tanner_PRB_2019}
\bibinfo{author}{\bibfnamefont{D.~S.~P.} \bibnamefont{Tanner}},
  \bibinfo{author}{\bibfnamefont{M.~A.} \bibnamefont{Caro}},
  \bibinfo{author}{\bibfnamefont{S.}~\bibnamefont{Schulz}}, \bibnamefont{and}
  \bibinfo{author}{\bibfnamefont{E.~P.} \bibnamefont{O'Reilly}},
  \bibinfo{journal}{Phys.~Rev.~B} \textbf{\bibinfo{volume}{100}},
  \bibinfo{pages}{094112} (\bibinfo{year}{2019}{\natexlab{b}}).

\bibitem[{\citenamefont{Gale}(1997)}]{Gale_JCSFT_1997}
\bibinfo{author}{\bibfnamefont{J.~D.} \bibnamefont{Gale}},
  \bibinfo{journal}{JCS Faraday Trans.} \textbf{\bibinfo{volume}{93}},
  \bibinfo{pages}{629} (\bibinfo{year}{1997}).

\bibitem[{\citenamefont{Gale and Rohl}(2003)}]{Gale_MS_2003}
\bibinfo{author}{\bibfnamefont{J.~D.} \bibnamefont{Gale}} \bibnamefont{and}
  \bibinfo{author}{\bibfnamefont{A.~L.} \bibnamefont{Rohl}},
  \bibinfo{journal}{Mol. Simul.} \textbf{\bibinfo{volume}{29}},
  \bibinfo{pages}{291} (\bibinfo{year}{2003}).

\bibitem[{\citenamefont{Gale}(2005)}]{Gale_ZK_2005}
\bibinfo{author}{\bibfnamefont{J.~D.} \bibnamefont{Gale}},
  \bibinfo{journal}{Z.~Krist.} \textbf{\bibinfo{volume}{220}},
  \bibinfo{pages}{552} (\bibinfo{year}{2005}).

\bibitem[{\citenamefont{O'Reilly and Lindsay}(2010)}]{Reilly_JPCS_2010}
\bibinfo{author}{\bibfnamefont{E.~P.} \bibnamefont{O'Reilly}} \bibnamefont{and}
  \bibinfo{author}{\bibfnamefont{A.}~\bibnamefont{Lindsay}},
  \bibinfo{journal}{J.~Phys.~Conf.~Ser.} \textbf{\bibinfo{volume}{242}},
  \bibinfo{pages}{012002} (\bibinfo{year}{2010}).

\bibitem[{\citenamefont{Vogl et~al.}(1983)\citenamefont{Vogl, Hjalmarson, and
  Dow}}]{Vogl_JPCS_1983}
\bibinfo{author}{\bibfnamefont{P.}~\bibnamefont{Vogl}},
  \bibinfo{author}{\bibfnamefont{H.~P.} \bibnamefont{Hjalmarson}},
  \bibnamefont{and} \bibinfo{author}{\bibfnamefont{J.~D.} \bibnamefont{Dow}},
  \bibinfo{journal}{J.~Phys.~Chem.~Solids} \textbf{\bibinfo{volume}{44}},
  \bibinfo{pages}{365} (\bibinfo{year}{1983}).

\bibitem[{\citenamefont{O'Halloran et~al.}(2019)\citenamefont{O'Halloran,
  Broderick, Tanner, Schulz, and O'Reilly}}]{Halloran_OQE_2019}
\bibinfo{author}{\bibfnamefont{E.~J.} \bibnamefont{O'Halloran}},
  \bibinfo{author}{\bibfnamefont{C.~A.} \bibnamefont{Broderick}},
  \bibinfo{author}{\bibfnamefont{D.~S.~P.} \bibnamefont{Tanner}},
  \bibinfo{author}{\bibfnamefont{S.}~\bibnamefont{Schulz}}, \bibnamefont{and}
  \bibinfo{author}{\bibfnamefont{E.~P.} \bibnamefont{O'Reilly}},
  \bibinfo{journal}{Opt.~Quant.~Electron.} \textbf{\bibinfo{volume}{51}},
  \bibinfo{pages}{314} (\bibinfo{year}{2019}).

\bibitem[{\citenamefont{Dresselhaus et~al.}(2008)\citenamefont{Dresselhaus,
  Dresselhaus, and Jorio}}]{Dresselhaus_book_2006}
\bibinfo{author}{\bibfnamefont{M.~S.} \bibnamefont{Dresselhaus}},
  \bibinfo{author}{\bibfnamefont{G.}~\bibnamefont{Dresselhaus}},
  \bibnamefont{and} \bibinfo{author}{\bibfnamefont{A.}~\bibnamefont{Jorio}},
  \emph{\bibinfo{title}{{Group Theory: Application to the Physics of Condensed
  Matter}}} (\bibinfo{publisher}{Springer}, \bibinfo{year}{2008}).

\bibitem[{\citenamefont{Jancu et~al.}(1998)\citenamefont{Jancu, Scholz,
  Beltram, and Bassani}}]{Jancu_PRB_1998}
\bibinfo{author}{\bibfnamefont{J.-M.} \bibnamefont{Jancu}},
  \bibinfo{author}{\bibfnamefont{R.}~\bibnamefont{Scholz}},
  \bibinfo{author}{\bibfnamefont{F.}~\bibnamefont{Beltram}}, \bibnamefont{and}
  \bibinfo{author}{\bibfnamefont{F.}~\bibnamefont{Bassani}},
  \bibinfo{journal}{Phys.~Rev.~B} \textbf{\bibinfo{volume}{57}},
  \bibinfo{pages}{6493} (\bibinfo{year}{1998}).

\bibitem[{\citenamefont{Lindsay and O'Reilly}(1999)}]{Lindsay_SSC_1999}
\bibinfo{author}{\bibfnamefont{A.}~\bibnamefont{Lindsay}} \bibnamefont{and}
  \bibinfo{author}{\bibfnamefont{E.~P.} \bibnamefont{O'Reilly}},
  \bibinfo{journal}{Solid State Commun.} \textbf{\bibinfo{volume}{112}},
  \bibinfo{pages}{443} (\bibinfo{year}{1999}).

\bibitem[{\citenamefont{O'Reilly et~al.}(2002)\citenamefont{O'Reilly, Lindsay,
  Tomi\'{c}, and Kamal-Saadi}}]{Reilly_SST_2002}
\bibinfo{author}{\bibfnamefont{E.~P.} \bibnamefont{O'Reilly}},
  \bibinfo{author}{\bibfnamefont{A.}~\bibnamefont{Lindsay}},
  \bibinfo{author}{\bibfnamefont{S.}~\bibnamefont{Tomi\'{c}}},
  \bibnamefont{and}
  \bibinfo{author}{\bibfnamefont{M.}~\bibnamefont{Kamal-Saadi}},
  \bibinfo{journal}{Semicond.~Sci.~Technol.} \textbf{\bibinfo{volume}{17}},
  \bibinfo{pages}{870} (\bibinfo{year}{2002}).

\bibitem[{\citenamefont{Lindsay and
  O'Reilly}(2004{\natexlab{a}})}]{Lindsay_PRL_2004}
\bibinfo{author}{\bibfnamefont{A.}~\bibnamefont{Lindsay}} \bibnamefont{and}
  \bibinfo{author}{\bibfnamefont{E.~P.} \bibnamefont{O'Reilly}},
  \bibinfo{journal}{Phys.~Rev.~Lett.} \textbf{\bibinfo{volume}{93}},
  \bibinfo{pages}{196402} (\bibinfo{year}{2004}{\natexlab{a}}).

\bibitem[{\citenamefont{Lindsay and O'Reilly}(2007)}]{Lindsay_PRB_2007}
\bibinfo{author}{\bibfnamefont{A.}~\bibnamefont{Lindsay}} \bibnamefont{and}
  \bibinfo{author}{\bibfnamefont{E.~P.} \bibnamefont{O'Reilly}},
  \bibinfo{journal}{Phys.~Rev.~B} \textbf{\bibinfo{volume}{76}},
  \bibinfo{pages}{075210} (\bibinfo{year}{2007}).

\bibitem[{\citenamefont{Lindsay and O'Reilly}(2008)}]{Lindsay_PSSC_2008}
\bibinfo{author}{\bibfnamefont{A.}~\bibnamefont{Lindsay}} \bibnamefont{and}
  \bibinfo{author}{\bibfnamefont{E.~P.} \bibnamefont{O'Reilly}},
  \bibinfo{journal}{Phys. Status Solidi C} \textbf{\bibinfo{volume}{5}},
  \bibinfo{pages}{454} (\bibinfo{year}{2008}).

\bibitem[{\citenamefont{Sander et~al.}(2011)\citenamefont{Sander, Teubert,
  Klar, Lindsay, and O'Reilly}}]{Sander_PRB_2011}
\bibinfo{author}{\bibfnamefont{T.}~\bibnamefont{Sander}},
  \bibinfo{author}{\bibfnamefont{J.}~\bibnamefont{Teubert}},
  \bibinfo{author}{\bibfnamefont{P.~J.} \bibnamefont{Klar}},
  \bibinfo{author}{\bibfnamefont{A.}~\bibnamefont{Lindsay}}, \bibnamefont{and}
  \bibinfo{author}{\bibfnamefont{E.~P.} \bibnamefont{O'Reilly}},
  \bibinfo{journal}{Phys.~Rev.B} \textbf{\bibinfo{volume}{83}},
  \bibinfo{pages}{235213} (\bibinfo{year}{2011}).

\bibitem[{\citenamefont{Usman et~al.}(2011)\citenamefont{Usman, Broderick,
  Lindsay, and O'Reilly}}]{Usman_PRB_2011}
\bibinfo{author}{\bibfnamefont{M.}~\bibnamefont{Usman}},
  \bibinfo{author}{\bibfnamefont{C.~A.} \bibnamefont{Broderick}},
  \bibinfo{author}{\bibfnamefont{A.}~\bibnamefont{Lindsay}}, \bibnamefont{and}
  \bibinfo{author}{\bibfnamefont{E.~P.} \bibnamefont{O'Reilly}},
  \bibinfo{journal}{Phys.~Rev.~B} \textbf{\bibinfo{volume}{84}},
  \bibinfo{pages}{245202} (\bibinfo{year}{2011}).

\bibitem[{\citenamefont{Usman et~al.}(2013)\citenamefont{Usman, Broderick,
  Batool, Hild, Hosea, Sweeney, and O'Reilly}}]{Usman_PRB_2013}
\bibinfo{author}{\bibfnamefont{M.}~\bibnamefont{Usman}},
  \bibinfo{author}{\bibfnamefont{C.~A.} \bibnamefont{Broderick}},
  \bibinfo{author}{\bibfnamefont{Z.}~\bibnamefont{Batool}},
  \bibinfo{author}{\bibfnamefont{K.}~\bibnamefont{Hild}},
  \bibinfo{author}{\bibfnamefont{T.~J.~C.} \bibnamefont{Hosea}},
  \bibinfo{author}{\bibfnamefont{S.~J.} \bibnamefont{Sweeney}},
  \bibnamefont{and} \bibinfo{author}{\bibfnamefont{E.~P.}
  \bibnamefont{O'Reilly}}, \bibinfo{journal}{Phys.~Rev.~B}
  \textbf{\bibinfo{volume}{87}}, \bibinfo{pages}{115104}
  (\bibinfo{year}{2013}).

\bibitem[{\citenamefont{Broderick et~al.}(2014)\citenamefont{Broderick,
  Mazzucato, Carr\`{e}re, Amand, Makhloufi, Arnoult, Fontaine, Donmez, Erol,
  Usman et~al.}}]{Broderick_PRB_2014}
\bibinfo{author}{\bibfnamefont{C.~A.} \bibnamefont{Broderick}},
  \bibinfo{author}{\bibfnamefont{S.}~\bibnamefont{Mazzucato}},
  \bibinfo{author}{\bibfnamefont{H.}~\bibnamefont{Carr\`{e}re}},
  \bibinfo{author}{\bibfnamefont{T.}~\bibnamefont{Amand}},
  \bibinfo{author}{\bibfnamefont{H.}~\bibnamefont{Makhloufi}},
  \bibinfo{author}{\bibfnamefont{A.}~\bibnamefont{Arnoult}},
  \bibinfo{author}{\bibfnamefont{C.}~\bibnamefont{Fontaine}},
  \bibinfo{author}{\bibfnamefont{O.}~\bibnamefont{Donmez}},
  \bibinfo{author}{\bibfnamefont{A.}~\bibnamefont{Erol}},
  \bibinfo{author}{\bibfnamefont{M.}~\bibnamefont{Usman}},
  \bibnamefont{et~al.}, \bibinfo{journal}{Phys.~Rev.~B}
  \textbf{\bibinfo{volume}{90}}, \bibinfo{pages}{195301}
  (\bibinfo{year}{2014}).

\bibitem[{\citenamefont{Usman et~al.}(2018)\citenamefont{Usman, Broderick, and
  O'Reilly}}]{Usman_PRA_2018}
\bibinfo{author}{\bibfnamefont{M.}~\bibnamefont{Usman}},
  \bibinfo{author}{\bibfnamefont{C.~A.} \bibnamefont{Broderick}},
  \bibnamefont{and} \bibinfo{author}{\bibfnamefont{E.~P.}
  \bibnamefont{O'Reilly}}, \bibinfo{journal}{Phys.~Rev.~Applied}
  \textbf{\bibinfo{volume}{10}}, \bibinfo{pages}{044024}
  (\bibinfo{year}{2018}).

\bibitem[{\citenamefont{Brey et~al.}(1984)\citenamefont{Brey, Tejedor, and
  Verg\'{e}s}}]{Brey_PRB_1984}
\bibinfo{author}{\bibfnamefont{L.}~\bibnamefont{Brey}},
  \bibinfo{author}{\bibfnamefont{C.}~\bibnamefont{Tejedor}}, \bibnamefont{and}
  \bibinfo{author}{\bibfnamefont{J.~A.} \bibnamefont{Verg\'{e}s}},
  \bibinfo{journal}{Phys.~Rev.~B} \textbf{\bibinfo{volume}{29}},
  \bibinfo{pages}{6840} (\bibinfo{year}{1984}).

\bibitem[{\citenamefont{Priester et~al.}(1988)\citenamefont{Priester, Allan,
  and Lannoo}}]{Priester_PRB_1988}
\bibinfo{author}{\bibfnamefont{C.}~\bibnamefont{Priester}},
  \bibinfo{author}{\bibfnamefont{G.}~\bibnamefont{Allan}}, \bibnamefont{and}
  \bibinfo{author}{\bibfnamefont{M.}~\bibnamefont{Lannoo}},
  \bibinfo{journal}{Phys.~Rev.~B} \textbf{\bibinfo{volume}{37}},
  \bibinfo{pages}{8519(R)} (\bibinfo{year}{1988}).

\bibitem[{\citenamefont{Slater and Koster}(1954)}]{Slater_PR_1954}
\bibinfo{author}{\bibfnamefont{J.~C.} \bibnamefont{Slater}} \bibnamefont{and}
  \bibinfo{author}{\bibfnamefont{G.~F.} \bibnamefont{Koster}},
  \bibinfo{journal}{Phys.~Rev.} \textbf{\bibinfo{volume}{94}},
  \bibinfo{pages}{1498} (\bibinfo{year}{1954}).

\bibitem[{\citenamefont{R\"{u}cker et~al.}(1983)\citenamefont{R\"{u}cker,
  Enderlein, and Bechstedt}}]{Rucker_PSSB_1983}
\bibinfo{author}{\bibfnamefont{H.}~\bibnamefont{R\"{u}cker}},
  \bibinfo{author}{\bibfnamefont{R.}~\bibnamefont{Enderlein}},
  \bibnamefont{and}
  \bibinfo{author}{\bibfnamefont{F.}~\bibnamefont{Bechstedt}},
  \bibinfo{journal}{Phys.~Status Solidi B} \textbf{\bibinfo{volume}{158}},
  \bibinfo{pages}{595} (\bibinfo{year}{1983}).

\bibitem[{\citenamefont{Mu{\~{n}}oz and Armelles}(1993)}]{Munoz_PRB_1993}
\bibinfo{author}{\bibfnamefont{M.~C.} \bibnamefont{Mu{\~{n}}oz}}
  \bibnamefont{and} \bibinfo{author}{\bibfnamefont{G.}~\bibnamefont{Armelles}},
  \bibinfo{journal}{Phys.~Rev.~B} \textbf{\bibinfo{volume}{48}},
  \bibinfo{pages}{2839} (\bibinfo{year}{1993}).

\bibitem[{\citenamefont{Li et~al.}(2009)\citenamefont{Li, Walsh, Chen, Yin,
  Yang, Li, Silva, Gong, and Wei}}]{Li_APL_2009}
\bibinfo{author}{\bibfnamefont{Y.-H.} \bibnamefont{Li}},
  \bibinfo{author}{\bibfnamefont{A.}~\bibnamefont{Walsh}},
  \bibinfo{author}{\bibfnamefont{S.}~\bibnamefont{Chen}},
  \bibinfo{author}{\bibfnamefont{W.-J.} \bibnamefont{Yin}},
  \bibinfo{author}{\bibfnamefont{J.-H.} \bibnamefont{Yang}},
  \bibinfo{author}{\bibfnamefont{J.}~\bibnamefont{Li}},
  \bibinfo{author}{\bibfnamefont{J.~L.~F.~D.} \bibnamefont{Silva}},
  \bibinfo{author}{\bibfnamefont{X.~G.} \bibnamefont{Gong}}, \bibnamefont{and}
  \bibinfo{author}{\bibfnamefont{S.-H.} \bibnamefont{Wei}},
  \bibinfo{journal}{Appl.~Phys.~Lett.} \textbf{\bibinfo{volume}{94}},
  \bibinfo{pages}{212109} (\bibinfo{year}{2009}).

\bibitem[{\citenamefont{Kirwan et~al.}(2018)\citenamefont{Kirwan, O'Halloran,
  Broderick, Schulz, and O'Reilly}}]{Kirwan_IEEENano_2018}
\bibinfo{author}{\bibfnamefont{A.~C.} \bibnamefont{Kirwan}},
  \bibinfo{author}{\bibfnamefont{E.~J.} \bibnamefont{O'Halloran}},
  \bibinfo{author}{\bibfnamefont{C.~A.} \bibnamefont{Broderick}},
  \bibinfo{author}{\bibfnamefont{S.}~\bibnamefont{Schulz}}, \bibnamefont{and}
  \bibinfo{author}{\bibfnamefont{E.~P.} \bibnamefont{O'Reilly}},
  \bibinfo{journal}{Proc.~18$^{\textrm{th}}$ IEEE International Conference on
  Nanotechnology}  (\bibinfo{year}{2018}).

\bibitem[{\citenamefont{Eales et~al.}(2019)\citenamefont{Eales, Marko, Schulz,
  O'Halloran, Ghetmiri, Du, Zhou, Yu, Margetis, Tolle et~al.}}]{Eales_SR_2019}
\bibinfo{author}{\bibfnamefont{T.~D.} \bibnamefont{Eales}},
  \bibinfo{author}{\bibfnamefont{I.~P.} \bibnamefont{Marko}},
  \bibinfo{author}{\bibfnamefont{S.}~\bibnamefont{Schulz}},
  \bibinfo{author}{\bibfnamefont{E.~J.} \bibnamefont{O'Halloran}},
  \bibinfo{author}{\bibfnamefont{S.}~\bibnamefont{Ghetmiri}},
  \bibinfo{author}{\bibfnamefont{W.}~\bibnamefont{Du}},
  \bibinfo{author}{\bibfnamefont{Y.}~\bibnamefont{Zhou}},
  \bibinfo{author}{\bibfnamefont{S.-Q.} \bibnamefont{Yu}},
  \bibinfo{author}{\bibfnamefont{J.}~\bibnamefont{Margetis}},
  \bibinfo{author}{\bibfnamefont{J.}~\bibnamefont{Tolle}},
  \bibnamefont{et~al.}, \bibinfo{journal}{Sci.~Rep.}
  \textbf{\bibinfo{volume}{9}}, \bibinfo{pages}{14077} (\bibinfo{year}{2019}).

\bibitem[{\citenamefont{Broderick
  et~al.}(2019{\natexlab{c}})\citenamefont{Broderick, O'Halloran, and
  O'Reilly}}]{Broderick_NUSOD_2019}
\bibinfo{author}{\bibfnamefont{C.~A.} \bibnamefont{Broderick}},
  \bibinfo{author}{\bibfnamefont{E.~J.} \bibnamefont{O'Halloran}},
  \bibnamefont{and} \bibinfo{author}{\bibfnamefont{E.~P.}
  \bibnamefont{O'Reilly}}, \bibinfo{journal}{Proc.~19$^{\textrm{th}}$
  International Conference on Numerical Simulation of Optoelectronic Devices}
  p. \bibinfo{pages}{117} (\bibinfo{year}{2019}{\natexlab{c}}).

\bibitem[{\citenamefont{Thouless}(1974)}]{Thouless_PR_1974}
\bibinfo{author}{\bibfnamefont{D.~J.} \bibnamefont{Thouless}},
  \bibinfo{journal}{Phys.~Rep.} \textbf{\bibinfo{volume}{13}},
  \bibinfo{pages}{93} (\bibinfo{year}{1974}).

\bibitem[{\citenamefont{Tanner et~al.}(2016)\citenamefont{Tanner, Caro,
  O'Reilly, and Schulz}}]{Tanner_RSCA_2016}
\bibinfo{author}{\bibfnamefont{D.~S.~P.} \bibnamefont{Tanner}},
  \bibinfo{author}{\bibfnamefont{M.~A.} \bibnamefont{Caro}},
  \bibinfo{author}{\bibfnamefont{E.~P.} \bibnamefont{O'Reilly}},
  \bibnamefont{and} \bibinfo{author}{\bibfnamefont{S.}~\bibnamefont{Schulz}},
  \bibinfo{journal}{RSC Adv.} \textbf{\bibinfo{volume}{6}},
  \bibinfo{pages}{64513} (\bibinfo{year}{2016}).

\bibitem[{\citenamefont{Lindsay and
  O'Reilly}(2004{\natexlab{b}})}]{Lindsay_PE_2004}
\bibinfo{author}{\bibfnamefont{A.}~\bibnamefont{Lindsay}} \bibnamefont{and}
  \bibinfo{author}{\bibfnamefont{E.~P.} \bibnamefont{O'Reilly}},
  \bibinfo{journal}{Physica E} \textbf{\bibinfo{volume}{21}},
  \bibinfo{pages}{901} (\bibinfo{year}{2004}{\natexlab{b}}).

\bibitem[{\citenamefont{Harris et~al.}(2008)\citenamefont{Harris, Lindsay, and
  O'Reilly}}]{Harris_JPCM_2008}
\bibinfo{author}{\bibfnamefont{C.}~\bibnamefont{Harris}},
  \bibinfo{author}{\bibfnamefont{A.}~\bibnamefont{Lindsay}}, \bibnamefont{and}
  \bibinfo{author}{\bibfnamefont{E.~P.} \bibnamefont{O'Reilly}},
  \bibinfo{journal}{J.~Phys.: Condens.~Matter} \textbf{\bibinfo{volume}{20}},
  \bibinfo{pages}{295211} (\bibinfo{year}{2008}).

\bibitem[{\citenamefont{Zhang and Wang}(2011)}]{Zhang_PRB_2011}
\bibinfo{author}{\bibfnamefont{Y.}~\bibnamefont{Zhang}} \bibnamefont{and}
  \bibinfo{author}{\bibfnamefont{L.-W.} \bibnamefont{Wang}},
  \bibinfo{journal}{Phys.~Rev.~B} \textbf{\bibinfo{volume}{83}},
  \bibinfo{pages}{165208} (\bibinfo{year}{2011}).

\bibitem[{\citenamefont{Kent et~al.}(2002)\citenamefont{Kent, Bellaiche, and
  Zunger}}]{Kent_SST_2002}
\bibinfo{author}{\bibfnamefont{P.~R.~C.} \bibnamefont{Kent}},
  \bibinfo{author}{\bibfnamefont{L.}~\bibnamefont{Bellaiche}},
  \bibnamefont{and} \bibinfo{author}{\bibfnamefont{A.}~\bibnamefont{Zunger}},
  \bibinfo{journal}{Semicond.~Sci.~Technol.} \textbf{\bibinfo{volume}{17}},
  \bibinfo{pages}{851} (\bibinfo{year}{2002}).

\bibitem[{\citenamefont{Broderick
  et~al.}(2017{\natexlab{a}})\citenamefont{Broderick, Seifikar, O'Reilly, and
  Rorison}}]{Broderick_nitride_chapter_2017}
\bibinfo{author}{\bibfnamefont{C.~A.} \bibnamefont{Broderick}},
  \bibinfo{author}{\bibfnamefont{M.}~\bibnamefont{Seifikar}},
  \bibinfo{author}{\bibfnamefont{E.~P.} \bibnamefont{O'Reilly}},
  \bibnamefont{and} \bibinfo{author}{\bibfnamefont{J.~M.}
  \bibnamefont{Rorison}}, \emph{\bibinfo{title}{Chapter 9: Dilute Nitride
  Alloys, Handbook of Optoelectronic Device Modeling and Simulation, Vol.~1}}
  (\bibinfo{publisher}{CRC Press}, \bibinfo{year}{2017}{\natexlab{a}}).

\bibitem[{\citenamefont{Broderick
  et~al.}(2017{\natexlab{b}})\citenamefont{Broderick, Marko, O'Reilly, and
  Sweeney}}]{Broderick_bismide_chapter_2017}
\bibinfo{author}{\bibfnamefont{C.~A.} \bibnamefont{Broderick}},
  \bibinfo{author}{\bibfnamefont{I.~P.} \bibnamefont{Marko}},
  \bibinfo{author}{\bibfnamefont{E.~P.} \bibnamefont{O'Reilly}},
  \bibnamefont{and} \bibinfo{author}{\bibfnamefont{S.~J.}
  \bibnamefont{Sweeney}}, \emph{\bibinfo{title}{Chapter 10: Dilute Bismide
  Alloys, Handbook of Optoelectronic Device Modeling and Simulation, Vol.~1}}
  (\bibinfo{publisher}{CRC Press}, \bibinfo{year}{2017}{\natexlab{b}}).

\bibitem[{\citenamefont{Kent and Zunger}(2001)}]{Kent_PRB_2001}
\bibinfo{author}{\bibfnamefont{P.~R.~C.} \bibnamefont{Kent}} \bibnamefont{and}
  \bibinfo{author}{\bibfnamefont{A.}~\bibnamefont{Zunger}},
  \bibinfo{journal}{Phys.~Rev.~B} \textbf{\bibinfo{volume}{64}},
  \bibinfo{pages}{115208} (\bibinfo{year}{2001}).

\end{thebibliography}

\end{document}